\documentclass[showpacs,aps,prd,nofootinbib,floatfix,amsmath,amssymb]{revtex4}
\usepackage{graphicx}
\usepackage[tight,TABTOPCAP]{subfigure}
\usepackage{color}
\begin{document}

\makeatletter
%Feynman slash
\newbox\slashbox \setbox\slashbox=\hbox{$/$}
\newbox\Slashbox \setbox\Slashbox=\hbox{\large$/$}
\def\pFMslash#1{\setbox\@tempboxa=\hbox{$#1$}
  \@tempdima=0.5\wd\slashbox \advance\@tempdima 0.5\wd\@tempboxa
  \copy\slashbox \kern-\@tempdima \box\@tempboxa}
\def\pFMSlash#1{\setbox\@tempboxa=\hbox{$#1$}
  \@tempdima=0.5\wd\Slashbox \advance\@tempdima 0.5\wd\@tempboxa
  \copy\Slashbox \kern-\@tempdima \box\@tempboxa}
\def\FMslash{\protect\pFMslash}
\def\FMSlash{\protect\pFMSlash}
\def\miss#1{\ifmmode{/\mkern-11mu #1}\else{${/\mkern-11mu #1}$}\fi}
%%%% Uso:  \pFMSlash{p}

\def\Nequal#1%
{\mathrel{\raisebox{-.5em}{$\stackrel{=}{\scriptstyle\rm#1}$}}}

\makeatother

%\tightenlines
\title{Lorentz-violating effects on pair production of $W$ bosons in photon collisions}
\author{J. I. Aranda$^{(a)}$, F. Ram\'\i rez-Zavaleta$^{(a)}$, F. J. Tlachino$^{(b)}$, J. J. Toscano$^{(b)}$, E. S. Tututi$^{(a)}$}
\address{
$^{(a)}$Facultad de Ciencias F\'\i sico Matem\' aticas,
Universidad Michoacana de San Nicol\' as de
Hidalgo, Avenida Francisco J. M\' ujica S/N, 58060, Morelia, Michoac\'an, M\' exico. \\
$^{(b)}$Facultad de Ciencias F\'{\i}sico Matem\'aticas,
Benem\'erita Universidad Aut\'onoma de Puebla, Apartado Postal
1152, Puebla, Puebla, M\'exico.}
\begin{abstract}
We examine Lorentz-violating effects that could appear through deviations of the Standard Model gauge couplings $WW\gamma$ and $WW\gamma\gamma$. These new physics effects are explored on the $\gamma\gamma \to WW$ reaction at the International Linear Collider. In particular, the associated helicity amplitudes are computed in the context of the Standard Model Extension (which is a model-independent approach to $CPT$ and Lorentz violation) and the Effective Lagrangian Model (which incorporates new physics effects that respect $CPT$ and Lorentz
violation). We perform an exhaustive study of the polarized differential cross sections to stand out effects related to Lorentz symmetry violation, where it is evidenced that the effects of Lorentz symmetry violation are more sensitive to the presence of the $\mathbf{b}$ constant background field. We found that for the $(\pm,\pm,(L,T+T,L))$ polarization state, only Standard Model Extension and Effective Lagrangian Model contribute at the lowest order, however, both types of new physics effects are clearly distinguished, being dominant the convoluted cross section of the Standard Model Extension in around 4 orders of magnitude. For this polarization state, at the last stage of operation of International Linear Collider, it is expected an integrated luminosity of $10^3$ fb$^{-1}$, finding around of 2 events for a Lorentz-violating energy scale of $32$ TeV.
\end{abstract}


\maketitle
\section{Introduction}
\label{i}
The International Linear Collider (ILC) is an ambitious project of electron-positron, electron-photon, and photon-photon collisions at the TeV energy scale~\cite{ILC, ILC-2013}, which will offers a clean environment to make studies beyond the capabilities of the Large Hadron Collider (LHC). Moreover, the operation of this collider in the $\gamma\gamma$ mode provides an excellent opportunity to explore new physics effects through production mechanisms that are not accessible in the hadronic machine. The photon collider might disclose crucial information on those production mechanisms that are naturally suppressed in electron-positron collisions, since the involved cross sections can be significantly larger than the corresponding $e^+e^-$ ones. These could be the case of new physics effects related to Lorentz violation (LV), which are absent within the Standard Model (SM). This type of new physics effect can induce deviations on observables that are sensitive to spatial orientation. Although these effects may be insignificant if the observable in consideration is computed taking into account all spatial directions, they could show up with strong intensity in some preferred directions. This is the case of polarized cross sections associated with collision processes involving particles with nonzero spin, which usually are strongly depending on the scattering angle.

It has been suggested the possibility that Lorentz symmetry can be violated at very small distances or very high energies. For instance, it has predicted that certain mechanism in string theory~\cite{KS} or in quantum gravity~\cite{QG} can induce the violation of Lorentz symmetry. Since these theories have not been totally developed, an effective field theory that contains both the SM and gravity has been formulated. Moreover, exists a minimal version without gravity~\cite{SME}, which is called the Standard Model Extension (SME)~\cite{SMEG}. The SME provides us a powerful tool for investigating LV effects in a model-independent manner. LV is also a feature of field theories formulated in a noncommutative space-time~\cite{Snyder}. This type of theories have been received particular attention since Seiberg and Witten showed how to connect commutative and noncommutative gauge field theories~\cite{SWM}. A method to formulate the SM as an effective field theory (NCSM), which is expressed in powers of the noncommutativity parameter, has been proposed in Refs.~\cite{NCYM,NCSM}. In fact, as it has been shown in Ref.~\cite{SME-NCSM}, the NCSM is a subset of the SME. Although these effective theories introduce constant background fields that carry Lorentz indices, they are not Lorentz invariants under general Lorentz transformations, but only under observer Lorentz transformations. As it has been discussed in Refs.~\cite{SME-NCSM}, there are two distinct classes of Lorentz transformations, namely, the observer and particle Lorentz transformations. The former corresponds to a change of coordinates, whereas the latter can be associated with a change of the measurement apparatus~\cite{Rev}.

In this work, we are interested in determining the involved physical consequences on the $\gamma \gamma \to WW$ process due to the presence of a constant background field characterized by an antisymmetric tensor $b^{\alpha \beta}$. This background field can arise from quantum gravity with spontaneous symmetry breaking or from a noncommutative spacetime. This is an interesting reaction, which will be under the scrutiny of the ILC~\cite{ILC} operating in the $\gamma\gamma$ mode. Although the radiative corrections are important within the SM, for our purposes will be sufficient to compare our results with the SM prediction at the Born approximation. In the SM, the tree-level cross section for $\gamma \gamma\to WW$ has been previously studied~\cite{TSM, TSM1, RSM} and an exhaustive analysis of the one-loop radiative corrections was given in Ref.~\cite{RSM}. This process provides a good mechanism to investigate the presence of new physics effects on the $WW\gamma$ and $WW\gamma\gamma$ vertices. New physics effects on the $\gamma \gamma\to WW$ reaction has been studied by some authors~\cite{WW1,WW2,WW3,WW4,WW5,WW6,WW7,WW8,WW9,WW10,WW11,WW12,WW13} beyond the SM. Such a new physics has been traditionally considered through an effective vertex $WW\gamma$, which is parametrized by means of form factors that characterize the electromagnetic properties of the $W$ gauge boson. Gaemers and Gounaris~\cite{GG} derived initially 9 form factors for the $WW\gamma$ vertex, but further on a careful analysis carried out by Hagiwara-Peccei-Zeppenfeld-Hikasa~\cite{Hagiwara} showed that only 7 of these quantities are independent indeed.  These form factors define the charge, the magnetic and electric dipole moments, the magnetic and electric quadrupole moments, and the CP-even and CP-odd anapole moments of this particle. Although model independent, it is assumed that these form factors respect both the Lorentz symmetry and the SM gauge symmetry. In other words, the sources of new physics have nothing to do with LV. As already mentioned, in this work we are interested in finding deviations of the SM prediction for the $\gamma \gamma \to WW$ process by assuming the presence of a Lorentz violating effective $WW\gamma$ and $WW\gamma\gamma$ vertices, whose source may be, for instance, general relativity with spontaneous symmetry breaking or a noncommutative space-time. However, we will adopt a model-independent approach by using the general formalism of the SME~\cite{SME}. The structure of the effective Lagrangian characterizing the SME~\cite{SME,T1} differs substantially from that describing the Conventional Effective Standard Model (CESM)~\cite{EL} extension. While the SME is constructed out by gauge-invariant Lorentz tensors contracted with constants Lorentz tensors specifying preferred spatial directions, the CESM are made of objects that are both gauge invariant and Lorentz invariant or, equivalently, of gauge-invariant Lorentz tensors appropriately contracted with products of metric tensors. Thus, it is expected that the SM deviations induced by anomalous $WW\gamma$ and $WW\gamma\gamma$ vertices on the $\gamma\gamma\to WW$ process differs from one to other approach. An important goal of this work is to investigate not only the deviations of the SM prediction to the  $\gamma \gamma\to WW$ reaction due to Lorentz violating effects present in the $WW\gamma$ and $WW\gamma\gamma$ vertices, but also to compare these deviations with those induced by other sources of new physics effects parametrized in the scheme of CESM. This type of information will be valuable in future experiments. We will focus on the Yang-Mills part of the effective Lagrangian that characterizes the SME (or also the NCSM) modified by the presence of an observer invariant that arises from the contraction of the constant antisymmetric tensor  $b^{\alpha \beta}$ with a Lorentz 2-tensor that is invariant under the $SU_L(2)$ gauge group.  This extended Yang-Mills sector generates nonrenormalizable $WW\gamma$ and $WW\gamma\gamma$ vertices, which differs substantially from the ones studied in references~\cite{W1,W2,W3,W4,W5,W6} within the context of the CESM~\cite{EL}. In general, for each observer invariant constructed with a Lorentz $k$-tensor contracted with a $k$-tensor background field in the SME, there is a counterpart in the context of the CESM that results from the contraction of such Lorentz $k$-tensor with an appropriate product of the metric tensor. To simplify our analysis as much as possible, we will consider the simplest extension of the $SU_L(2)$ Yang-Mills sector in both the SME and the CESM. Explicit expressions for the helicity amplitudes of the $\gamma \gamma\to WW$ scattering including a detailed analysis of their angular distributions will be presented.

The rest of the paper has been organized as follows. In Sec. \ref{L}, effective Lagrangians for the Yang-Mills sector of the $SU_L(2)$ group that includes gauge-invariant interactions of up to dimension-six in both the CESM and the SME are presented. In particular, the main differences of the gauge and Lorentz structure of the $WW\gamma$ and $WW\gamma\gamma$ vertices arising from each of these effective formulations of new physics are discussed. In Sec.~\ref{EGIA}, we present SM, CESM, and SME amplitudes with explicit gauge-invariant structure. Sec. \ref{ha} is devoted to calculate the helicity amplitudes for the $\gamma \gamma\to WW$ reaction. In Sec. \ref{D}, we discuss our results. Finally, in Sec. \ref{C}, the final remarks are presented.

\section{Effective Yang-Mills Lagrangian }
\label{L}
The study of the gauge structure of the $WWV$ vertex ($V=\gamma,Z$) has been the subject of important works in different contexts. The one-loop radiative corrections to the renormalizable vertex have been calculated in the SM~\cite{Bardeen} and some of its extensions~\cite{BSM}. The radiative corrections to these vertices with the $\gamma$ and $Z$ bosons off shell have been studied in the SM using a linear gauge~\cite{SMLG} and also via the Pinch Technique~\cite{PT}. Virtual effects of new heavy gauge bosons to these off-shell vertices have been studied in a covariant way under the electroweak group within the context of $331$ models~\cite{T2} and in theories with universal extra dimensions~\cite{ED}. Its most general structure has been parametrized in a model independent manner in the context of CESM~\cite{GG,Hagiwara} and used in various phenomenological applications~\cite{W1,W2,W3,W4,W5,W6,Other}. As commented, the $WW\gamma$ effective vertex that arises from the CESM approach differs substantially from the one that can be constructed in the context of the SME.

The other important vertex for our discussion is the $WW\gamma\gamma$ coupling, which directly arises from the Yang-Mills Lagrangian. This vertex appears within the context of the SM and it has been the subject of important studies in the literature~\cite{WW1,WW2,WW4,WW5,WW6,WW13}. As we will see, the $WW\gamma\gamma$ vertex receives contributions from the effective anomalous Yang-Mills sectors, corresponding to the CESM~\cite{EL} and the SME~\cite{SME,T1}. In order to clarify details of the calculations, let us discuss with some extent how the $WW\gamma$ and the $WW\gamma\gamma$ couplings emerge in both the CESM and SME descriptions.

The building blocks needed to introduce $SU_L(2)$ and $U_Y(1)$ invariant operators of arbitrary dimension are the respective tensors $W_{\mu \nu}=T^aW^a_{\mu \nu}$ and $B_{\mu \nu}$. These gauge-invariant operators are all Lorentz tensors of even rank, which will be denoted as ${\cal O}_{\mu_1,\mu_2,\cdots \mu_{2n}}$. One can set up operators that are invariant under general Lorentz transformations by contracting these gauge-invariant operators with a tensor of the same rank made of a product of metric tensors, that is, $g^{\mu_1,\mu_2}\cdots g^{\mu_{2n-1}\mu_{2n}}{\cal O}_{\mu_1,\mu_2,\cdots \mu_{2n}}$ . Alternatively, one can construct Lorentz-observer invariant operators by contracting these gauge-invariant operators with a constant tensor of the same rank, that is,  $b^{\mu_1 \mu_2\cdots \mu_{2n}}{\cal O}_{\mu_1,\mu_2,\cdots \mu_{2n}}$. The former scheme leads us to the CESM, which is a technique that allows us to parametrize, in a model-independent fashion, effects of new physics that respect both the gauge symmetry and the Lorentz symmetry. On the other hand, when one adopts the latter scheme, one arrives to the SME~\cite{SME}, which is an effective field theory that allows to incorporate $CPT$ violation and Lorentz violation in a model-independent manner. It should be noticed that the gauge structure is the same in both CESM and SME approach to new physics. The fundamental difference between both schemes comes from the method in which the Lorentz-invariant action is constructed. In the CESM approach, Lorentz invariance is established through contractions with the metric tensor, which is a self-invariant Lorentz object by definition. In contrast, in the SME approach, a Lorentz-observer invariant action can be constructed by contracting the Lorentz $2n$-tensor operators with constant background $2n$-tensors which are true tensorial objects under Lorentz-observer transformations, but not under Lorentz-particle transformations~\cite{SME,Rev}. This general scheme comprises the very interesting situation in which the $b^{\mu_1 \mu_2\cdots \mu_{2n}}$ constant tensor corresponds to a vacuum expectation value of a tensor field $B^{\mu_1 \mu_2\cdots \mu_{2n}}(x)$. This particular case corresponds to a spontaneous symmetry breaking of the Lorentz group, which arises in specific scenarios of string theories of general relativity.

\begin{figure}[htb!]
\centering\includegraphics[width=1.5in]{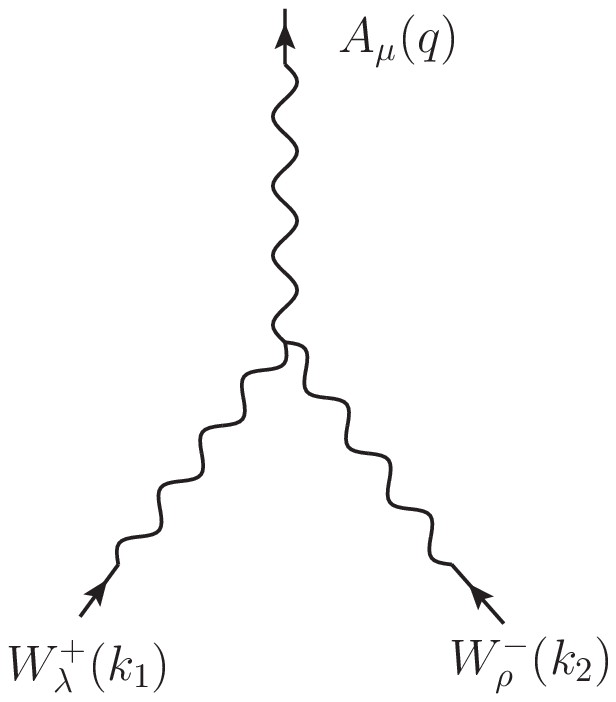}
\caption{\label{V}The trilinear $WW\gamma$ vertex.}
\end{figure}

As mentioned in the introduction, in this paper we will focus on the electroweak Yang-Mills sector of the SM. The gauge-invariant Lorentz tensor operators of up to dimension six that can be constructed with the $W_{\mu \nu}$ and $B_{\mu \nu}$ are the following\footnote{Other possible dimension-six Lorentz 2-tensor that can be constructed is $Tr[W_{\alpha \beta}W_{\mu \nu}W^{\mu \nu}]$, but it vanishes, as $ W^b_{\mu \nu}W^{c\mu \nu}Tr[\sigma^a \sigma^b \sigma^c]=2i W^b_{\mu \nu}W^{c\mu \nu}\epsilon^{abc}=0$.}:
\begin{eqnarray}
SU_L(2): &&  {\cal O}^W_{\alpha \beta \lambda \rho}=Tr[W_{\alpha \beta}W_{\lambda \rho}]\, , \, \, {\cal O}_{\alpha \beta}=Tr[W_{\alpha \lambda} W_{\beta \rho} W^{\lambda  \rho}] \, , \\
U_Y(1): && {\cal O}^B_{\alpha \beta \lambda \rho}=B_{\alpha \beta}B_{\lambda \rho}\, .
\end{eqnarray}
Within the context context of the CESM approach, the contractions $g^{\alpha \beta}g^{\lambda \rho}{\cal O}^W_{\alpha \beta \lambda \rho}$, $g^{\alpha \beta}g^{\lambda \rho}{\cal O}^B_{\alpha \beta \lambda \rho}$, and $g^{\alpha \beta}{\cal O}_{\alpha \beta}$ lead to an effective electroweak Yang-Mills sector that includes up to dimension six interactions, which can conveniently be written as
\begin{equation}
{\cal L}_{CESM}^{YM}=-\frac{1}{4}W^a_{\mu \nu}W^{\mu \nu}_a-\frac{1}{4}B_{\mu \nu}B^{\mu \nu}+\frac{g\, \alpha_W}{\Lambda^2}\frac{\epsilon_{abc}}{3!}W^a_{\alpha \lambda}W^{b \alpha}_{\, \, \, \rho}W^{c\lambda \rho}\, ,
\end{equation}
where some constant factors have been introduced. In particular, $\Lambda$ represents the new physics scale and $\alpha_W$ is an unknown coefficient, which can be calculated once the fundamental theory is known. To write down the most general Lorentz structure of the $WW\gamma$ vertex it is necessary to introduce additional dimension-six operators whose construction involves a Higgs doublet~\cite{EL,GG,Hagiwara}, but for simplicity we do not consider them. It should be noticed that due to the symmetric character of the metric tensor, only the symmetric part of the ${\cal O}_{\alpha \beta}$ operator contributes to the $WW\gamma$ coupling. In this context, the  $WW\gamma$ vertex of the CESM is given by the Lagrangian
\begin{equation}
{\cal L}_{WW\gamma}^{CESM} = {\cal L}^{SM}_{WW\gamma}+{\cal L}^{\alpha_W}_{WW\gamma},
\end{equation}
where
\begin{equation}
{\cal L}^{\alpha_{W}}_{WW\gamma}=\frac{ie\alpha_W}{\Lambda^2}W^{-}_{\lambda \rho}W^{+\lambda}_{\, \, \eta}F^{\rho \eta},
\end{equation}
and the corresponding vertex function can be expressed, in the unitary gauge, as
\begin{equation}
\Gamma^{CESM}_{\mu\lambda\rho} =\Gamma^{SM}_{\mu\lambda\rho} + \Gamma^{\alpha_W}_{\mu\lambda\rho},
\end{equation}
where
\begin{align}
\Gamma^{SM}_{\mu\lambda\rho}(q,k_1,k_2) &= i e\left[(k_1 - k_2)_{\mu} g_{\lambda\rho} + (q - k_1 )_{\rho} g_{\lambda\mu} - (q - k_2 )_{\lambda} g_{\rho\mu}\right],\\
\Gamma^{\alpha_{W}}_{\mu\lambda \rho}(q,k_1,k_2)&=\frac{ie\alpha_W}{\Lambda^2}\left(q^\eta\delta^\beta_\mu-q^\beta \delta^\eta_\mu \right)\left(k^\alpha_1g_{\eta \lambda}-k_{1\eta}\delta^\alpha_\lambda \right)\left(k_{2\alpha}g_{\beta \rho}-k_{2\beta}g_{\alpha \rho} \right)\, .
\end{align}
$\Gamma^{SM}_{\mu\lambda\rho}(q,k_1,k_2)$ and $\Gamma^{\alpha_{W}}_{\mu\lambda \rho }(q,k_1,k_2)$ represent SM and pure anomalous contributions, respectively. We have employed the notation and conventions shown in Fig. \ref{V}. Notice that this vertex satisfies the following simple Ward identities
\begin{eqnarray}
q^\mu \Gamma^{\alpha_{W}}_{\lambda \rho \mu}(q,k_1,k_2)&=&0\, ,\\
k^\lambda_2\Gamma^{\alpha_{W}}_{\lambda \rho \mu}(q,k_1,k_2)&=&0\, ,\\
k^\rho _3\Gamma^{\alpha_{W}}_{\lambda \rho \mu}(q,k_1,k_2)&=&0\, .
\end{eqnarray}

On the other hand, following a similar treatment to that used to determine the $WW\gamma$ coupling, we can obtain the Lagrangian associated with the $WW\gamma\gamma$ vertex, which is given by
\begin{equation}
{\cal L}_{WW\gamma\gamma}^{CESM} = {\cal L}^{SM}_{WW\gamma\gamma}+{\cal L}^{\alpha_W}_{WW\gamma\gamma},
\end{equation}
where ${\cal L}^{SM}_{WW\gamma\gamma}$ is the SM contribution, which is well known. The anomalous contribution is given by
\begin{equation}
{\cal L}^{\alpha_W}_{WW\gamma\gamma} = \frac{e^{2}\alpha_{W}}{\Lambda^{2}}F^{\rho\eta}[W^{-}_{\lambda\rho}(A_{\eta}W^{+\lambda} - W^{+}_{\eta}A^{\lambda}) + (A_{\lambda}W^{-}_{\rho} - W^{-}_{\lambda}A_{\rho})W^{+\lambda}_{\eta}].
\end{equation}

\begin{figure}[htb!]
\centering\includegraphics[width=1.75in]{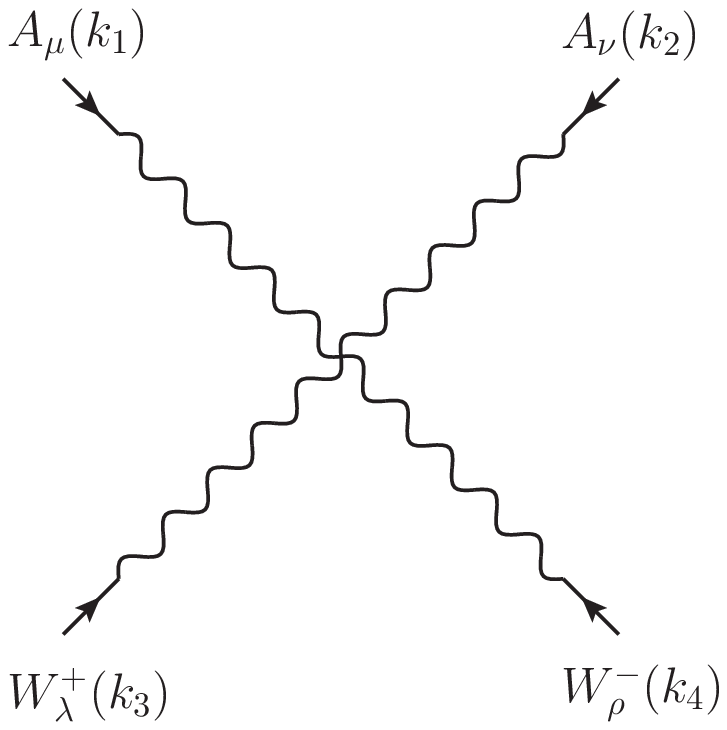}
\caption{\label{vwwff}The $WW\gamma\gamma$ vertex.}
\end{figure}

From ${\cal L}_{WW\gamma\gamma}^{CESM}$, it follows that the vertex function relating to the $WW\gamma\gamma$ coupling~\footnote{The notation of Fig.~\ref{vwwff} must be used to properly interpret the Feynman rule.} can be written as
\begin{equation}
\Gamma^{CESM}_{\mu\nu\lambda\rho} =\Gamma^{SM}_{\mu\nu\lambda\rho} + \Gamma^{\alpha_W}_{\mu\nu\lambda\rho},
\end{equation}
where
\begin{equation}
\Gamma^{SM}_{\mu\nu\lambda\rho} = -i e^2 \left[2g_{\lambda\rho}g_{\mu\nu} - (g_{\lambda\mu}g_{\rho\nu}+g_{\rho\mu}g_{\lambda\nu})\right],
\end{equation}
is the SM vertex function and
\begin{equation}
\Gamma^{\alpha_W}_{\mu\nu\lambda\rho}(k_1,k_2,k_3,k_4) = \frac{ie^{2}\alpha_{W}}{\Lambda^{2}}(\Gamma^{\sigma\eta}_{\mu}\Gamma_{\sigma\eta\lambda\rho\nu}(k_3,k_4) + \Gamma^{\sigma\eta}_{\nu}\Gamma_{\sigma\eta\lambda\rho\mu}(k_3,k_4)),
\end{equation}
is the anomalous vertex function, being
\begin{eqnarray}
\Gamma^{\sigma\eta}_{\mu} &=& k^{\sigma}_{1}\delta^{\eta}_{\mu}-k^{\eta}_{1}\delta^{\sigma}_{\mu},\\
\Gamma^{\sigma\eta}_{\nu} &=& k^{\sigma}_{2}\delta^{\eta}_{\nu}-k^{\eta}_{2}\delta^{\sigma}_{\nu},
\end{eqnarray}
with
\begin{equation}
\Gamma_{\sigma\eta\lambda\rho\nu}(k_3,k_4) = (g_{\alpha\nu}g_{\eta\lambda}-g_{\eta\nu}g_{\alpha\lambda})(k_{4\alpha}g_{\sigma\rho}-k_{4\sigma}g_{\alpha\rho})
- (g_{\alpha\nu}g_{\sigma\rho}-g_{\sigma\nu}g_{\alpha\rho})(k_{3\alpha}g_{\eta\lambda}-k_{3\eta}g_{\alpha\lambda}).
\end{equation}
$\Gamma_{\sigma\eta\lambda\rho\mu}(k_3,k_4)$ is derived from $\Gamma_{\sigma\eta\lambda\rho\nu}(k_3,k_4)$ by replacing $\nu$ with $\mu$.

%The anomalous vertex function $\Gamma^{\alpha_W}_{\mu\nu\lambda\rho}$ satisfies the following identities
%\begin{align}
%k_{1}^{\mu}\Gamma^{\alpha_W}_{\mu\nu\lambda\rho} &= \frac{ie^{2}\alpha_{W}}{\Lambda^{2}}\Gamma^{\sigma\eta}_{\nu}k_{1}^{\mu}\Gamma_{\sigma\eta\lambda\rho\mu}\\
%k_{2}^{\nu}\Gamma^{\alpha_W}_{\mu\nu\lambda\rho} &= \frac{ie^{2}\alpha_{W}}{\Lambda^{2}}\Gamma^{\sigma\eta}_{\mu}k_{2}^{\nu}\Gamma_{\sigma\eta\lambda\rho\nu}\\
%k_{1}^{\mu}k_{2}^{\nu}\Gamma^{\alpha_W}_{\mu\nu\lambda\rho} &= 0.
%\end{align}

After discussing the structure of the $WW\gamma$ and $WW\gamma\gamma$ vertices within the context of CESM, we proceed to introduce an effective Yang-Mills Lagrangian that generates these vertices in the context of the SME. The main differences between the CESM and the SME, as well as the fact that the NCSM is a subset of the SME~\cite{SME-NCSM}, have been discussed with some extent in reference~\cite{T1}. Here, we will only present those features that are important for our purposes. As already commented, we will consider a minimal scenario in which no degrees of freedom different of the gauge fields associated with the electroweak group are considered. In addition, our scenario will be one in which constant background tensors couples with gauge tensors, but not with their dual ones. Within the framework of the SME, a Lorentz-observer invariant and $CPT$-conserving effective electroweak Yang-Mills sector can be constructed by contracting the ${\cal O}^W_{\alpha \beta \lambda \rho}$, ${\cal O}^B_{\alpha \beta \lambda \rho}$, and ${\cal O}_{\alpha \beta}$ operators with Lorentz constant tensors. The corresponding Lagrangian can be written as follows:
\begin{equation}
{\cal L}^{YM}_{SME}=-\frac{1}{4}W^a_{\mu \nu}W^{\mu \nu}_a-\frac{1}{4}B_{\mu \nu}B^{\mu \nu}+k^{\alpha \beta \lambda \rho}_W{\cal O}^W_{\alpha \beta \lambda \rho}+k^{\alpha \beta \lambda \rho}_B{\cal O}^B_{\alpha \beta \lambda \rho}+b^{\alpha \beta}{\cal O}_{\alpha \beta} \, ,
\end{equation}
where the SM Yang-Mills sector has been included. The dimensionless constant tensors $k^{\alpha \beta \lambda \rho}_{W,B}$ are antisymmetric under the interchanges $\alpha \leftrightarrow \beta$ and $\lambda \leftrightarrow \rho$, but are symmetric under the simultaneous interchange of the pairs of indices $(\alpha \beta) \leftrightarrow (\lambda \rho)$~\cite{SME}. Moreover, it is assumed that the constant 2-tensor $b^{\alpha \beta}$, which has units of squared mass, is antisymmetric. This means that only the antisymmetric part of the ${\cal O}_{\alpha \beta}$ operator contributes to the SME, in contrast with the CESM approach, in which only the symmetric part of this operator contributes. This is a crucial feature that allows us to distinguish one approach from the other. In particular, as we will see below, the symmetric or antisymmetric anomalous contributions of ${\cal O}_{\alpha \beta}$ to the $WW\gamma$ and $WW\gamma\gamma$ vertices will be reflected in the helicity amplitudes for the $\gamma \gamma \to WW$ scattering. As it has been discussed in the context of string theory quantization~\cite{SWM} and in general relativity with spontaneous symmetry breaking~\cite{SSBQG}, there exists more than a simple analogy between the six $b^{\alpha \beta}$ quantities and the six components of the electromagnetic field tensor $F^{\alpha \beta}$. These six independent components, given by $e^i\equiv \Lambda^2_{LV} b^{0i}$ and $b^i\equiv (1/2)\Lambda^2_{LV}\epsilon^{ijk}b^{jk}$, with $\Lambda_{LV}$ the new physics scale, determine two preferred spatial directions, which play the role of an external agent that would induce deviations from the SM predictions which in principle could be observed in future high-energy experiments.

Regarding the renormalizable operators ${\cal O}^W_{\alpha \beta \lambda \rho}$ and  ${\cal O}^B_{\alpha \beta \lambda \rho}$, they have already been considered in the literature in other contexts. Besides to modify the $WW\gamma$ vertex, these operators also introduce changes in the photon propagator and through it induce contributions to some cosmological observables, which impose severe constraints on these class of operators~\cite{KM,KR}. Due to this, in this work we will not consider Lorentz-violating effects coming from these renormalizable interactions. Then, the anomalous contribution to the $WW\gamma$ vertex in the context of the SME only arises from the nonrenormalizable term $b^{\alpha \beta}{\cal O}_{\alpha \beta}$. This contribution is given by
\begin{equation}
\mathcal{L}^{NC}_{WW\gamma}= \frac{ie}{2}b^{\alpha
\beta}(W^-_{\alpha \lambda}W^+_{\beta \rho}F^{\lambda
\rho}+W^+_{\alpha \lambda}W^{-\lambda \rho}F_{\beta
\rho}+W^-_{\beta \rho}W^{+\lambda \rho}F_{\alpha \lambda}).
\end{equation}
Using the notation and conventions shown in Fig.\ref{V}, the corresponding vertex function can be written as follows:
\begin{equation}
\Gamma_{\mu \lambda \rho}(q,k_1,k_2)=\frac{ie}{2}\,b^{\alpha \beta}T^{\eta \xi}_{\mu}\Gamma_{\alpha \beta \eta \xi \lambda \rho}\, \, ,
\end{equation}
where
\begin{equation}
T^{\eta \xi}_{\mu}=q^{\xi}\delta^{\eta}_{\mu}-q^{\eta}\delta^{\xi}_{\mu},
\end{equation}
and
\begin{align}
\Gamma_{\alpha \beta \eta \xi \lambda \rho}(k_1,k_2) & = (k_{1 \beta}g_{\xi \lambda}-k_{1 \xi}g_{\beta \lambda})(k_{2 \alpha}g_{\eta \rho}-k_{2 \eta}g_{\alpha \rho})
+g_{\eta \beta}(k_{1 \alpha}g_{\sigma \lambda}-k_{1 \sigma}g_{\alpha \lambda})(k_2^{\sigma}g_{\xi \rho}-k_{2 \xi}\delta^{\sigma}_{\rho}) {}\nonumber \\
& {} +g_{\eta \alpha}(k_{1 \xi}\delta^{\sigma}_{\lambda}-k_{1 \sigma}g_{\xi \lambda})(k_{2 \beta}g_{\sigma \rho}-k_{2 \sigma}g_{\beta \rho}) \, \, .
\end{align}
From this expression, it is evident that $\Gamma_{
\mu\lambda \rho}(q,k_1,k_2)$  satisfies the following simple Ward
identities~\cite{Tla}:
\begin{eqnarray}
q^\mu \Gamma_{\mu\lambda \rho}(q,k_1,k_2)=0, \\
k^\lambda_1\Gamma_{\mu\lambda \rho}(q,k_1,k_2)=0, \\
k^\rho_2\Gamma_{\mu\lambda \rho}(q,k_1,k_2)=0.
\end{eqnarray}

On the other hand, following a similar way to that used in the study of the anomalous $WW\gamma$ contribution, we find that the effect of the background field $b^{\alpha\beta}$ on the $WW\gamma\gamma$ vertex is represented by the following Lagrangian:
\begin{eqnarray}\label{lwwgg}
{\cal L}^{NC}_{WW\gamma\gamma} &=& -\frac{e^{2}}{2}b^{\alpha\beta}F^{\sigma\eta}[W^{-}_{\alpha\sigma}(A_{\beta}W^{+}_{\eta}-A_{\eta}W^{+}_{\beta}) - W^{+}_{\beta\eta}(A_{\alpha}W^{-}_{\sigma}-A_{\sigma}W^{-}_{\alpha})\nonumber\\
 & & + g_{\beta\sigma}(W^{-}_{\omega\eta}(A_{\alpha}W^{+\omega}-A^{\omega}W^{+}_{\alpha}) - W^{+}_{\alpha\omega}(A^{\omega}W^{-}_{\eta}-A_{\eta}W^{-\omega}))\nonumber\\
  & & + g_{\alpha\sigma}(W^{-}_{\beta\omega}(A_{\eta}W^{+\omega}-A^{\omega}W^{+}_{\eta}) - W^{+}_{\eta\omega}(A_{\beta}W^{-\omega}-A^{\omega}W^{-}_{\omega}))].
\end{eqnarray}
From Eq.~(\ref{lwwgg}), we can obtain the anomalous vertex function for the $WW\gamma\gamma$ coupling, which can be written as follows (see Fig.~\ref{vwwff})
\begin{equation}
\Gamma^{b}_{\mu\nu\lambda\rho} = -\frac{ie^2}{2}b^{\alpha\beta}(\Gamma^{\sigma\eta}_{\mu}\Gamma_{\alpha\beta\sigma\eta\lambda\rho\nu} + \Gamma^{\sigma\eta}_{\nu}\Gamma_{\alpha\beta\sigma\eta\lambda\rho\mu}),
\end{equation}
where
\begin{align}
\Gamma^{\sigma\eta}_{\mu} &= k^{\sigma}_{1}\delta^{\eta}_{\mu}-k^{\eta}_{1}\delta^{\sigma}_{\mu},\\
\Gamma^{\sigma\eta}_{\nu} &= k^{\sigma}_{2}\delta^{\eta}_{\nu}-k^{\eta}_{2}\delta^{\sigma}_{\nu},
\end{align}
with
\begin{align}
\Gamma_{\alpha\beta\sigma\eta\lambda\rho\nu}(k_3,k_4) &= (g_{\beta\nu}g_{\eta\lambda}-g_{\eta\nu}g_{\beta\lambda})(k_{4\alpha}g_{\sigma\rho}-k_{4\sigma}g_{\alpha\rho})
- (g_{\alpha\nu}g_{\sigma\rho}-g_{\sigma\nu}g_{\alpha\rho})(k_{3\beta}g_{\eta\lambda}-k_{3\eta}g_{\beta\lambda})\nonumber\\
& + g_{\beta\sigma}[g_{\alpha\nu}(k_{4\lambda}g_{\eta\rho}-k_{4\eta}g_{\lambda\rho}) - g_{\alpha\lambda}(k_{4\nu}g_{\eta\rho}-k_{4\eta}g_{\nu\rho})\nonumber\\
& - g_{\eta\rho}(k_{3\alpha}g_{\lambda\nu}-k_{3\nu}g_{\alpha\lambda}) - g_{\eta\nu}(k_{3\alpha}g_{\lambda\rho}-k_{3\rho}g_{\alpha\lambda})]\nonumber\\
& + g_{\alpha\sigma}[g_{\eta\nu}(k_{4\beta}g_{\lambda\rho}-k_{4\lambda}g_{\beta\rho}) - g_{\eta\lambda}(k_{4\beta}g_{\rho\nu}-k_{4\nu}g_{\beta\rho})\nonumber\\
& - g_{\beta\nu}(k_{3\eta}g_{\lambda\rho}-k_{3\rho}g_{\eta\lambda}) - g_{\beta\rho}(k_{3\eta}g_{\lambda\nu}-k_{3\nu}g_{\eta\lambda})].
\end{align}
Notice that $\Gamma_{\alpha\beta\sigma\eta\lambda\rho\mu}(k_3,k_4)$ can be constructed from $\Gamma_{\alpha\beta\sigma\eta\lambda\rho\nu}(k_3,k_4)$ by substituting $\nu$ with $\mu$.

%The vertex function $\Gamma^{b}_{\mu\nu\lambda\rho}$ satisfies the following identities
%\begin{align}
%k_{1}^{\mu}\Gamma^{b}_{\mu\nu\lambda\rho} &= -\frac{1}{2}b^{\alpha\beta}\Gamma^{\sigma\eta}_{\nu}k_{1}^{\mu}\Gamma_{\alpha\beta\sigma\eta\lambda\rho\mu}\nonumber\\
%k_{2}^{\nu}\Gamma^{b}_{\mu\nu\lambda\rho} &= -\frac{1}{2}b^{\alpha\beta}\Gamma^{\sigma\eta}_{\mu}k_{2}^{\nu}\Gamma_{\alpha\beta\sigma\eta\lambda\rho\nu}\nonumber\\
%k_{1}^{\mu}k_{2}^{\nu}\Gamma^{b}_{\mu\nu\lambda\rho} &= 0 \nonumber.
%\end{align}
%Clearly, we can see that
%\begin{eqnarray}
%k_{1}^{\mu}\Gamma_{\alpha\beta\sigma\eta\lambda\rho\mu}\neq 0\nonumber\\
%k_{2}^{\nu}\Gamma_{\alpha\beta\sigma\eta\lambda\rho\nu}\neq 0\nonumber.
%\end{eqnarray}

\section{The $\gamma \gamma\to WW$ process} \label{ha}
We now turn to calculate the helicity amplitudes for the $\gamma \gamma\to WW$ process mediated by the anomalous $WW\gamma$ and $WW\gamma\gamma$ vertices that arise in both the CESM and the SME. We will present our results in the center of mass reference frame.

\begin{figure}[htb!]
\centering
\includegraphics[height=2.5in,width=3.8in]{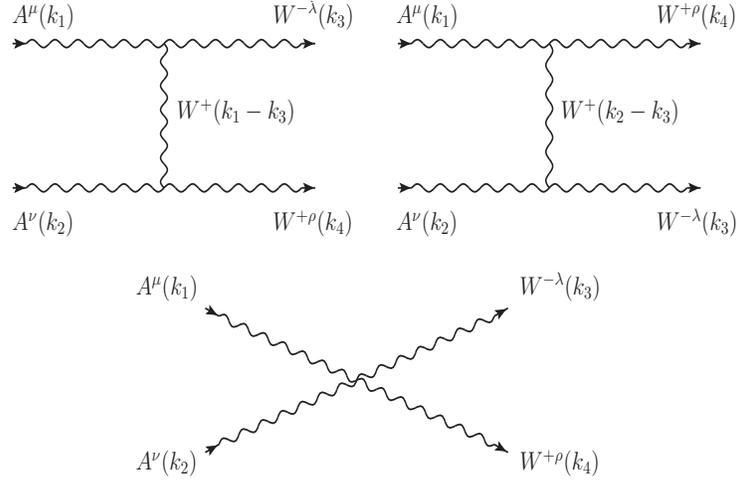}
\caption{\label{dfggww} Feynman diagrams contributing to the $\gamma\gamma\rightarrow WW$ reaction at the lowest order.}
\end{figure}

Our notation and conventions for the kinematics involved in the $\gamma \gamma\to WW$ reaction are shown in Figs.~\ref{dfggww} and \ref{cinematica}. The Lorentz indices and momenta are denoted as follows:
\begin{equation}
A^{\mu}(k_1)A^{\nu}(k_2)\to W^{-\lambda}(k_3)W^{+\rho}(k_4)
\end{equation}
\begin{eqnarray}
k_{1}^{\mu} &=& \frac{\sqrt{s}}{2}(1,0,0,1),\\
k_{2}^{\nu} &=& \frac{\sqrt{s}}{2}(1,0,0,-1),\\
k_{3}^{\lambda} &=& \frac{\sqrt{s}}{2}(1,\beta\sin\theta,0,\beta\cos\theta),\\
k_{4}^{\rho} &=& \frac{\sqrt{s}}{2}(1,-\beta\sin\theta,0,-\beta\cos\theta),
\end{eqnarray}
\begin{eqnarray}
\epsilon^{\mu}(k_1,\lambda_1) &=& \frac{1}{\sqrt{2}}(0,1,i\lambda_1,0),\\
\epsilon^{\nu}(k_2,\lambda_2) &=& \frac{1}{\sqrt{2}}(0,-1,i\lambda_2,0),\\
\epsilon^{*\lambda}(k_3,\lambda_3) &=& \frac{1}{\sqrt{2}}(0,\cos\theta,i\lambda_3,-\sin\theta),\\
\epsilon^{*\rho}(k_4,\lambda_4) &=& \frac{1}{\sqrt{2}}(0,-\cos\theta,i\lambda_4,\sin\theta),\\
\epsilon^{*\lambda}(k_3,\lambda_{3}^{0}) &=& \frac{\sqrt{s}}{2m_W}(\beta,\sin\theta,0,\cos\theta),\\
\epsilon^{*\rho}(k_4,\lambda_{4}^{0}) &=& \frac{\sqrt{s}}{2m_W}(\beta,-\sin\theta,0,-\cos\theta),
\end{eqnarray}
where $\beta=\sqrt{1-4m_{W}^{2}/s}$\,\,, $\lambda_i=\pm1$ for $i=1,2,3,4$ and $\lambda^{0}_{3,4}=0$. Here, $\theta$ symbolizes the scattering angle. Otherwise, Mandelstam variables are determined by
\begin{eqnarray}
s &=& (k_1 + k_2)^{2},\\
t &=& (k_1 - k_3)^{2}=-\frac{s}{2}\left( 1-\frac{2m^{2}_{W}}{s}-\beta\cos\theta \right),\\
t &=& (k_1 - k_4)^{2}=-\frac{s}{2}\left( 1-\frac{2m^{2}_{W}}{s}+\beta\cos\theta \right).
\end{eqnarray}

When considering all of the information forth above we can compute the polarized differential cross section in terms of helicity amplitudes. Thus, the polarized differential cross section can be written as
\begin{equation}
\left(\frac{d\sigma_{\lambda_1 \lambda_2 \overline{\lambda}_3 \overline{\lambda}_4}}{d\Omega}\right)_{CM} = \frac{1}{64\pi^{2}}\frac{\sqrt{s-4m_{W}^2}}{s^{3/2}}\left| {\cal M}_{\lambda_1\lambda_2\overline{\lambda}_3\overline{\lambda}_4} \right|^2 ,
\end{equation}
where ${\cal M}_{\lambda_1\lambda_2\overline{\lambda}_3\overline{\lambda}_4}$ are the helicity amplitudes, $\overline{\lambda}_i \equiv \lambda_{i}^{0},\lambda_{i}$ are the longitudinal and transverse $W$ boson helicity components, respectively.

Helicity amplitudes are composed of two parts as seen below
\begin{equation}
{\cal M}_{\lambda_1\lambda_2\overline{\lambda}_3\overline{\lambda}_4}={\cal M}_{\lambda_1\lambda_2\overline{\lambda}_3\overline{\lambda}_4}^{SM}+{\cal M}_{\lambda_1\lambda_2\overline{\lambda}_3\overline{\lambda}_4}^{NP},
\end{equation}
where the superscripts $SM$ and $NP$ stand for the SM and pure new physics contributions, respectively. As already commented, effects of physics beyond the Fermi scale on the $WW\gamma$ and $WW\gamma\gamma$ vertices will be considered in two different model-independent schemes, namely, the CESM extension, which respects both the Lorentz and SM gauge symmetries, and the SME approach, which respect the SM gauge symmetry but violates the Lorentz one.

\section{Explicit gauge-invariant amplitudes}
\label{EGIA}
Before obtaining the helicity amplitudes, we will analyze its gauge structure in the context of the SM, the CESM and the SME.
The contributions to the $\gamma\gamma\rightarrow WW$ process are shown in Fig.~\ref{dfggww} and respective calculations were performed in the unitary gauge. We will present manifestly gauge-invariant amplitudes.

\subsection{The Standard Model gauge-invariant amplitude}

The SM model gauge-invariant amplitude can be written as
\begin{equation}
{\cal M}^{SM} = {\cal M}^{SM}_{\mu\nu\lambda\rho} \epsilon^{\mu}(k_{1},\lambda_{1}) \epsilon^{\nu}(k_{2},\lambda_{2}) \epsilon^{\lambda *}(k_{3},\overline{\lambda}_{3}) \epsilon^{\rho *}(k_{4},\overline{\lambda}_{4}).
\end{equation}
By properly grouping, it is possible to obtain an amplitude with explicit gauge invariance, which is given by
\begin{equation}
{\cal M}^{SM}_{\mu\nu\lambda\rho} = 2 i e ^{2} \sum^{5}_{i=1} N^{SM (i)}_{\mu\nu\lambda\rho},
\end{equation}
where
\begin{align}
N^{SM (1)}_{\mu\nu\lambda\rho} &= (k_{1\xi}g_{\lambda\mu} - k_{1\lambda}g_{\xi\mu})(k_{2\xi}g_{\rho\nu} - k_{2\rho}g_{\xi\nu}),\\
N^{SM (3)}_{\mu\nu\lambda\rho} &= \left(\frac{k_{4\mu}}{k_{1} \cdot k_{4}} - \frac{k_{3\mu}}{k_{1} \cdot k_{3}}\right)(k_{2\lambda}g_{\nu\rho} - k_{2\rho}g_{\lambda\nu}), \\
N^{SM (5)}_{\mu\nu\lambda\rho} &= g_{\lambda\rho}\left(\frac{k_{3\mu}k_{4\nu}}{k_{1} \cdot k_{3}} + \frac{k_{3\nu}k_{4\mu}}{k_{1} \cdot k_{4}} - g_{\mu\nu}\right) = g_{\lambda\rho}\left(\frac{k_{3\mu}k_{4\nu}}{k_{2} \cdot k_{4}} + \frac{k_{3\nu}k_{4\mu}}{k_{2} \cdot k_{3}} - g_{\mu\nu}\right).
\end{align}
The $N^{SM (2,4)}_{\mu\nu\lambda\rho}$ structures are obtained, respectively, from $N^{SM (1,3)}_{\mu\nu\lambda\rho}$ by invoking Bose symmetry. The $N^{i}_{\mu\nu\lambda\rho}$ structures are gauge invariant, i.e., satisfies simple Ward identities:

\begin{align}
\label{IW1}
N^{(i)}_{\mu\nu\lambda\rho}k^{\mu}_{1}&=0,\\
\label{IW2}
N^{(i)}_{\mu\nu\lambda\rho}k^{\nu}_{2}&=0.
\end{align}

\subsection{The CESM gauge-invariant amplitude}

The CESM gauge-invariant amplitude is given by
\begin{eqnarray}
{\cal M}^{CESM} &=& ({\cal M}^{SM}_{\mu\nu\lambda\rho} + {\cal M}^{\alpha_{W}}_{\mu\nu\lambda\rho}) \epsilon^{\mu}(k_{1},\lambda_{1}) \epsilon^{\nu}(k_{2},\lambda_{2}) \epsilon^{\lambda *}(k_{3},\overline{\lambda}_{3}) \epsilon^{\rho *}(k_{4},\overline{\lambda}_{4}),\\
              &=& {\cal M}^{SM} + {\cal M}^{\alpha_{W}}.
\end{eqnarray}
As in the previous section we analyzed the ${\cal M}^{SM}$ amplitude, we only focus on the anomalous contribution, to first order in $\alpha_{W}$, whose associated amplitude is given by
\begin{eqnarray}
-i{\cal M}^{\alpha_{W}}_{\mu\nu\lambda\rho} &=& \frac{\Gamma^{SM}_{\rho\chi\nu}(-k_{4},k_{1}-k_{3},k_{2}) {\Gamma^{\alpha_{W}}}^{\chi}_{\;\lambda\mu}(k_{3}-k_{1},-k_{3},k_{1})}{t-m^{2}_{W}}+ \frac{\Gamma^{SM}_{\chi\lambda\mu}(k_{3}-k_{1},-k_{3},k_{1}) {\Gamma^{\alpha_{W}}}^{\;\,\chi}_{\rho\;\nu}(-k_{4},k_{1}-k_{3},k_{2})}{t-m^{2}_{W}}\nonumber\\
&+& \frac{\Gamma^{SM}_{\rho\chi\mu}(-k_{4},k_{2}-k_{3},k_{1}) {\Gamma^{\alpha_{W}}}^{\chi}_{\;\lambda\nu}(k_{3}-k_{2},-k_{3},k_{2})}{u-m^{2}_{W}}+ \frac{\Gamma^{SM}_{\chi\lambda\nu}(k_{3}-k_{2},-k_{3},k_{2}) {\Gamma^{\alpha_{W}}}^{\;\,\chi}_{\rho\;\mu}(-k_{4},k_{2}-k_{3},k_{1})}{u-m^{2}_{W}}\nonumber\\
&+& \Gamma^{\alpha_{W}}_{\mu\nu\lambda\rho}(k_{1},k_{2},-k_{3},-k_{4}),
\end{eqnarray}
where $\Gamma^{SM,\,\alpha_{W}}_{\mu\nu\lambda}(k_{1},k_{2},k_{3})$ and $\Gamma^{SM,\,\alpha_W}_{\mu\nu\lambda\rho}(k_{1},k_{2},k_{3},k_{4})$ are the vertex functions  related to $WW \gamma$ and $ WW\gamma\gamma$ couplings, respectively. After performing algebraic manipulations, one can obtain an amplitude with explicit gauge invariance, which can be written as
\begin{equation}
{\cal M}^{\alpha_{W}}_{\mu\nu\lambda\rho} = \frac{i e^2 \alpha_W}{\Lambda^2} \sum^{10}_{i=1} N^{\alpha_W (i)}_{\mu\nu\lambda\rho},
\end{equation}
where
\begin{align}
N^{\alpha_W (1)}_{\mu\nu\lambda\rho} =& \frac{1}{k_2 \cdot k_4} (k^{\eta}_{1}\delta^{\beta}_{\mu}-k^{\beta}_{1}\delta^{\eta}_{\mu}) (k_{2\chi}g_{\rho\nu}-k_{2\rho}g_{\chi\nu}) (k_{3\beta}g_{\alpha\lambda}-k_{3\alpha}g_{\beta\lambda}) ((k_{2}-k_{4})^{\alpha}g_{\eta\chi}-(k_{2}-k_{4})_{\eta}\delta^{\alpha}_{\chi}),\\
N^{\alpha_W (3)}_{\mu\nu\lambda\rho} =& \frac{1}{k_2 \cdot k_4} (k_{1\lambda}g_{\chi\mu}-k_{1\chi}g_{\lambda\mu}) (k^{\eta}_{2}\delta^{\beta}_{\nu}-k^{\beta}_{2}\delta^{\eta}_{\nu}) (k_{4\eta}\delta^{\alpha}_{\rho}-k_{4}^{\alpha}g_{\eta\rho}) ((k_{2}-k_{3})_{\alpha}g_{\beta\chi}-(k_{1}-k_{3})_{\beta}g_{\alpha\chi}),\\
N^{\alpha_W (5)}_{\mu\nu\lambda\rho} =& (k^{\eta}_{1}\delta^{\beta}_{\mu}-k^{\beta}_{1}\delta^{\eta}_{\mu}) \left(\frac{k_{3\nu}}{k_{2} \cdot k_{3}} - \frac{k_{4\nu}}{k_{2} \cdot k_{4}}\right) (k_{3\beta}g_{\alpha\lambda}-k_{3\alpha}g_{\beta\lambda}) (k_{4\eta}\delta^{\alpha}_{\rho}-k_{4}^{\alpha}g_{\eta\rho}),\\
N^{\alpha_W (7)}_{\mu\nu\lambda\rho} =& (k^{\eta}_{1}\delta^{\beta}_{\mu}-k^{\beta}_{1}\delta^{\eta}_{\mu}) \left(\frac{k_{4\nu}}{k_{2} \cdot k_{4}} (k^{\alpha}_{2}g_{\eta\rho}-k_{2\eta}\delta^{\alpha}_{\rho}) - (\delta^{\alpha}_{\nu}g_{\eta\rho}-g_{\eta\nu}\delta^{\alpha}_{\rho})\right) (k_{3\beta}g_{\alpha\lambda}-k_{3\alpha}g_{\beta\lambda}),\\
N^{\alpha_W (9)}_{\mu\nu\lambda\rho} =& (k^{\eta}_{1}\delta^{\beta}_{\mu}-k^{\beta}_{1}\delta^{\eta}_{\mu}) \left(\frac{k_{3\nu}}{k_{2} \cdot k_{3}} (k_{2\alpha}g_{\beta\lambda}-k_{2\beta}g_{\alpha\lambda}) - (g_{\alpha\nu}g_{\beta\lambda}-g_{\beta\nu}g_{\alpha\lambda})\right) (k_{4\eta}\delta^{\alpha}_{\rho}-k_{4}^{\alpha}g_{\eta\rho}).
\end{align}
The remaining gauge structures are obtained by making use of the Bose symmetry. The $N^{\alpha_W (i)}_{\mu\nu\lambda\rho}$ structures are manifestly gauge invariant.

\subsection{The SME gauge-invariant amplitude}

The SME amplitude can be written as follows
\begin{align}
{\cal M}^{SME} &= ({\cal M}^{SM}_{\mu\nu\lambda\rho} + {\cal M}^{b}_{\mu\nu\lambda\rho}) \epsilon^{\mu}(k_{1},\lambda_{1}) \epsilon^{\nu}(k_{2},\lambda_{2}) \epsilon^{\lambda *}(k_{3},\overline{\lambda}_{3}) \epsilon^{\rho *}(k_{4},\overline{\lambda}_{4}),\\
              &= ({\cal M}^{SM} + {\cal M}^{b}).
\end{align}
As in the previous subsection, here, we only focus on the background field effect, to first order in $b^{\alpha\beta}$, whose associated amplitude is given by
\begin{eqnarray}
-i{\cal M}^{b}_{\mu\nu\lambda\rho} &=& \frac{\Gamma^{SM}_{\rho\chi\nu}(-k_{4},k_{1}-k_{3},k_{2}) {\Gamma^{b}}^{\chi}_{\;\,\lambda\mu}(k_{3}-k_{1},-k_{3},k_{1})}{t-m^{2}_{W}}+ \frac{\Gamma^{SM}_{\chi\lambda\mu}(k_{3}-k_{1},-k_{3},k_{1}) {\Gamma^{b}}^{\;\,\chi}_{\rho\;\nu}(-k_{4},k_{1}-k_{3},k_{2})}{t-m^{2}_{W}}\nonumber\\
& +& \frac{\Gamma^{SM}_{\rho\chi\mu}(-k_{4},k_{2}-k_{3},k_{1}) {\Gamma^{b}}^{\chi}_{\;\,\lambda\nu}(k_{3}-k_{2},-k_{3},k_{2})}{u-m^{2}_{W}}+ \frac{\Gamma^{SM}_{\chi\lambda\nu}(k_{3}-k_{2},-k_{3},k_{2}) {\Gamma^{b}}^{\;\,\chi}_{\rho\;\mu}(-k_{4},k_{2}-k_{3},k_{1})}{u-m^{2}_{W}}\nonumber\\
& +& \Gamma^{b}_{\mu\nu\lambda\rho}(k_{1},k_{2},-k_{3},-k_{4}),
\end{eqnarray}
where $\Gamma^{SM,\,b}_{\mu\nu\lambda}(k_{1},k_{2},k_{3})$ and $\Gamma^{SM,\,b}_{\mu\nu\lambda\rho}(k_{1},k_{2},k_{3},k_{4})$ are the vertex functions related to $WW \gamma$ and $ WW\gamma\gamma$ couplings, respectively. After performing algebraic manipulations, one can obtain an anomalous amplitude manifestly gauge invariance, which can be expressed as
\begin{equation}
{\cal M}^{b}_{\mu\nu\lambda\rho} = \frac{ie^2}{2} b^{\alpha\beta} \sum^{26}_{i=1} N^{b (i)}_{\alpha\beta\mu\nu\lambda\rho}.
\end{equation}
The explicit form of the $N^{b (i)}_{\alpha\beta\mu\nu\lambda\rho}$ gauge structures is given in the Appendix~\ref{APA}. It can easily be shown that structures $N^{b (i)}_{\alpha\beta\mu\nu\lambda\rho}$ satisfy simple Ward identities.

\section{Helicity Amplitudes}

\subsection{The Standard Model helicity amplitudes}
By performing contractions of the SM tensorial amplitude with photons and $W$ bosons polarization vectors, the corresponding helicity amplitudes can be written as follows
\begin{align}
{\cal M}^{SM}_{\lambda_1\lambda_2\lambda_3\lambda_4} = \frac{i e^{2}}{4 (\beta^2 \cos^2\theta - 1)} & \{-(\lambda_1 \lambda_2 + 3)(1 + \lambda_3 \lambda_4 ) \beta^2 + 4(\lambda_1 + \lambda_2)(\lambda_3 + \lambda_4)\beta \nonumber\\
          &- 6 \lambda_1 \lambda_2 \lambda_3 \lambda_4 + 2 \lambda_3 \lambda_4 - 4 + 4 (\lambda_1 - \lambda_2)(\lambda_3 - \lambda_4)\cos\theta \nonumber\\
          &+ (1 - \lambda_1 \lambda_2) [\beta^2 (1 + \lambda_3 \lambda_4 ) - 2] \cos(2\theta)\},
\end{align}
\begin{equation}
{\cal M}^{SM}_{\lambda_1\lambda_2\lambda^{0}_3\lambda^{0}_4} = \frac{i e^{2} s}{8 m^{2}_{W}( \beta^2 \cos^2\theta - 1)}(\beta^2 -1)\{-(\lambda_1 \lambda_2 + 3) \beta^2 + (\beta^2 -2) (1 - \lambda_1 \lambda_2) \cos(2 \theta) + 4\},
\end{equation}
\begin{equation}
{\cal M}^{SM}_{\lambda_1\lambda_2\lambda_3\lambda_4^{0}} = \frac{i e^{2}\sqrt{s}}{\sqrt{2} m_{W}( \beta^2 \cos^2\theta - 1)}(\beta^2 - 1)\{(1 - \lambda_1 \lambda_2) \cos\theta - (\lambda_1 - \lambda_2) \lambda_3\} \sin\theta,
\end{equation}
\begin{equation}
{\cal M}^{SM}_{\lambda_1\lambda_2\lambda_3^{0}\lambda_4} = \frac{i e^{2}\sqrt{s}}{\sqrt{2} m_{W}( \beta^2 \cos^2\theta - 1)}(\beta^2 - 1) \{(1 - \lambda_1 \lambda_2) \cos\theta +(\lambda_1 - \lambda_2 )\lambda_4\}\sin\theta.
\end{equation}
Bearing in mind that the different helicity states are organized as $(\lambda_1,\lambda_2,\lambda_3,\lambda_4)$, one easily can see that there are 36 helicity amplitudes of the SM, from which, 12 are exactly zero at this order of perturbation theory~\cite{RSM}: $(\pm,\pm,0,\pm)$, $(\pm,\pm,\pm,0)$, $(\pm,\pm,0,\mp)$, $(\pm,\pm,\mp,0)$, $(\pm,\pm,\pm,\mp)$, and $(\pm,\pm,\mp,\pm)$. Moreover, the following symmetries arise
\begin{align}
{\cal M}^{SM}_{\lambda_1\lambda_2\overline{\lambda}_3\overline{\lambda}_4}(s,t,u) &= {\cal M}^{SM}_{\lambda_2\lambda_1\overline{\lambda}_3\overline{\lambda}_4}(s,u,t),\\
{\cal M}^{SM}_{\lambda_1\lambda_2\overline{\lambda}_3\overline{\lambda}_4}(s,t,u) &= {\cal M}^{SM}_{\lambda_{-1}\lambda_{-2}\overline{\lambda}_{-4}\overline{\lambda}_{-3}}(s,u,t),\\
{\cal M}^{SM}_{\lambda_1\lambda_2\overline{\lambda}_3\overline{\lambda}_4}(s,t,u) &= {\cal M}^{SM}_{\lambda_{-2}\lambda_{-1}\overline{\lambda}_{-4}\overline{\lambda}_{-3}}(s,t,u),
\end{align}
which correspond to Bose, CP and Bose$+$CP symmetries~\cite{RSM}, respectively. In relation to parity and charge conjugation symmetries, we have that~\cite{RSM}
\begin{align}
{\cal M}^{SM}_{\lambda_1\lambda_2\lambda_3\lambda_4}(s,t,u)  & \Nequal{P}
{\cal M}^{SM}_{\lambda_{-1}\lambda_{-2}\lambda_{-3}\lambda_{-4}}(s,t,u),\\
{\cal M}^{SM}_{\lambda_1\lambda_2\lambda_3\lambda_4}(s,t,u)  & \Nequal{C}
{\cal M}^{SM}_{\lambda_{1}\lambda_{2}\lambda_{4}\lambda_{3}}(s,u,t).
\end{align}

\subsection{New physics effects in the CESM approach}
In the context of the CESM, the pure anomalous contribution is given as
\begin{align}
{\cal M}^{\alpha_{W}}_{\lambda_1\lambda_2\lambda_3\lambda_4} =& \frac{ie^{2}\alpha_{W}}{\Lambda^{2}}{\cal M}^{\alpha_{W}}_{\mu\nu\lambda\rho} \epsilon^{\mu}(k_{1},\lambda_{1}) \epsilon^{\nu}(k_{2},\lambda_{2}) \epsilon^{\lambda *}(k_{3},\lambda_{3}) \epsilon^{\rho *}(k_{4},\lambda_{4}),\\
{\cal M}^{\alpha_{W}}_{\lambda_1\lambda_2\lambda_3\lambda_{4}^{0}} =& \frac{ie^{2}\alpha_{W}}{\Lambda^{2}}{\cal M}^{\alpha_{W}}_{\mu\nu\lambda\rho} \epsilon^{\mu}(k_{1},\lambda_{1}) \epsilon^{\nu}(k_{2},\lambda_{2}) \epsilon^{\lambda *}(k_{3},\lambda_{3}) \epsilon^{\rho *}(k_{4},\lambda_{4}^{0}),\\
{\cal M}^{\alpha_{W}}_{\lambda_1\lambda_2\lambda_{3}^{0}\lambda_4} =& \frac{ie^{2}\alpha_{W}}{\Lambda^{2}}{\cal M}^{\alpha_{W}}_{\mu\nu\lambda\rho} \epsilon^{\mu}(k_{1},\lambda_{1}) \epsilon^{\nu}(k_{2},\lambda_{2}) \epsilon^{\lambda *}(k_{3},\lambda_{3}^{0}) \epsilon^{\rho *}(k_{4},\lambda_{4}),\\
{\cal M}^{\alpha_{W}}_{\lambda_1\lambda_2\lambda_{3}^{0}\lambda_{4}^{0}} =& \frac{ie^{2}\alpha_{W}}{\Lambda^{2}}{\cal M}^{\alpha_{W}}_{\mu\nu\lambda\rho} \epsilon^{\mu}(k_{1},\lambda_{1}) \epsilon^{\nu}(k_{2},\lambda_{2}) \epsilon^{\lambda *}(k_{3},\lambda_{3}^{0}) \epsilon^{\rho *}(k_{4},\lambda_{4}^{0}).
\end{align}
After some algebraic manipulations, the associated helicity amplitudes are derived
\begin{align}
{\cal M}^{\alpha_{W}}_{\lambda_1\lambda_2\lambda_3\lambda_4} &= \frac{ie^{2}\alpha_{W} s}{16 \Lambda^{2} (1 - \beta^2 \cos^2\theta)} \{ 3 (\lambda_1 + \lambda_2) (\lambda_3 + \lambda_4) \beta^3 - 2 (\lambda_1 \lambda_2 \lambda_3 \lambda_4 - \lambda_3 \lambda_4 - 2) \beta^2  \nonumber\\
& - 5 (\lambda_1 + \lambda_2) (\lambda_3 + \lambda_4) \beta - 2 (\lambda_1 \lambda_2 + 1) (2 \lambda_3 \lambda_4 + 1) \nonumber\\
& + [(\beta^3 + \beta) (\lambda_1 + \lambda_2) (\lambda_3 + \lambda_4) - \lambda_1 \lambda_2 (2 - (6 \lambda_3 \lambda_4 + 4) \beta^2) - 2] \cos(2\theta)\},\\
{\cal M}^{\alpha_{W}}_{\lambda_1\lambda_2\lambda^{0}_3\lambda^{0}_4} &= \frac{-ie^{2}\alpha_{W} s^{2}}{8 m_{W}^{2} \Lambda^{2} (\beta^2 \cos^2\theta - 1)}(\beta^2 - 1)^2 (\lambda_1 \lambda_2 + 1) \sin^2\theta,\\
{\cal M}^{\alpha_{W}}_{\lambda_1\lambda_2\lambda_3\lambda_4^{0}} &= \frac{-ie^{2}\alpha_{W} s^{3/2} (\beta^2 - 1) \sin\theta}{8\sqrt{2} m_{W} \Lambda^{2} (\beta^2 \cos^2\theta - 1)}\{(\lambda_1 - \lambda_2) \lambda_3 \beta^2 + [2 (\beta^2 - 1) \lambda_1 \lambda_2 + \beta (\lambda_1 + \lambda_2) \lambda_3 - 2] \cos\theta\},\\
{\cal M}^{\alpha_{W}}_{\lambda_1\lambda_2\lambda_3^{0}\lambda_4} &= \frac{-ie^{2}\alpha_{W} s^{3/2} (\beta^2 - 1) \sin\theta}{8\sqrt{2} m_{W} \Lambda^{2} (\beta^2 \cos^2\theta - 1)}\{(\lambda_2 - \lambda_1) \lambda_4 \beta^2 + [2 (\beta^2 - 1) \lambda_1 \lambda_2 + \beta (\lambda_1 + \lambda_2) \lambda_4 - 2] \cos\theta\}.
\end{align}
In these cases, it can be appreciated that helicity amplitudes with polarization states: $(\pm,\mp,\pm,\mp)$, $(\pm,\mp,\mp,\pm)$, and $(\pm,\mp,0,0)$, are exactly zero, however, these polarization states are different from zero in the context of the SM.

\subsection{New physics effects in the SME approach}
%%%%%%%%%%%%%%%%%%%%%%%%%%%%%%%%%%%%%%%%%%%%%%%%%%%%%%%%%%%%%%%%%%%%%%%%%%%%%%%%%%%%%%%%%%%%%%
\begin{figure}[htb!]
\centering
\includegraphics[height=2.5in,width=3.0in]{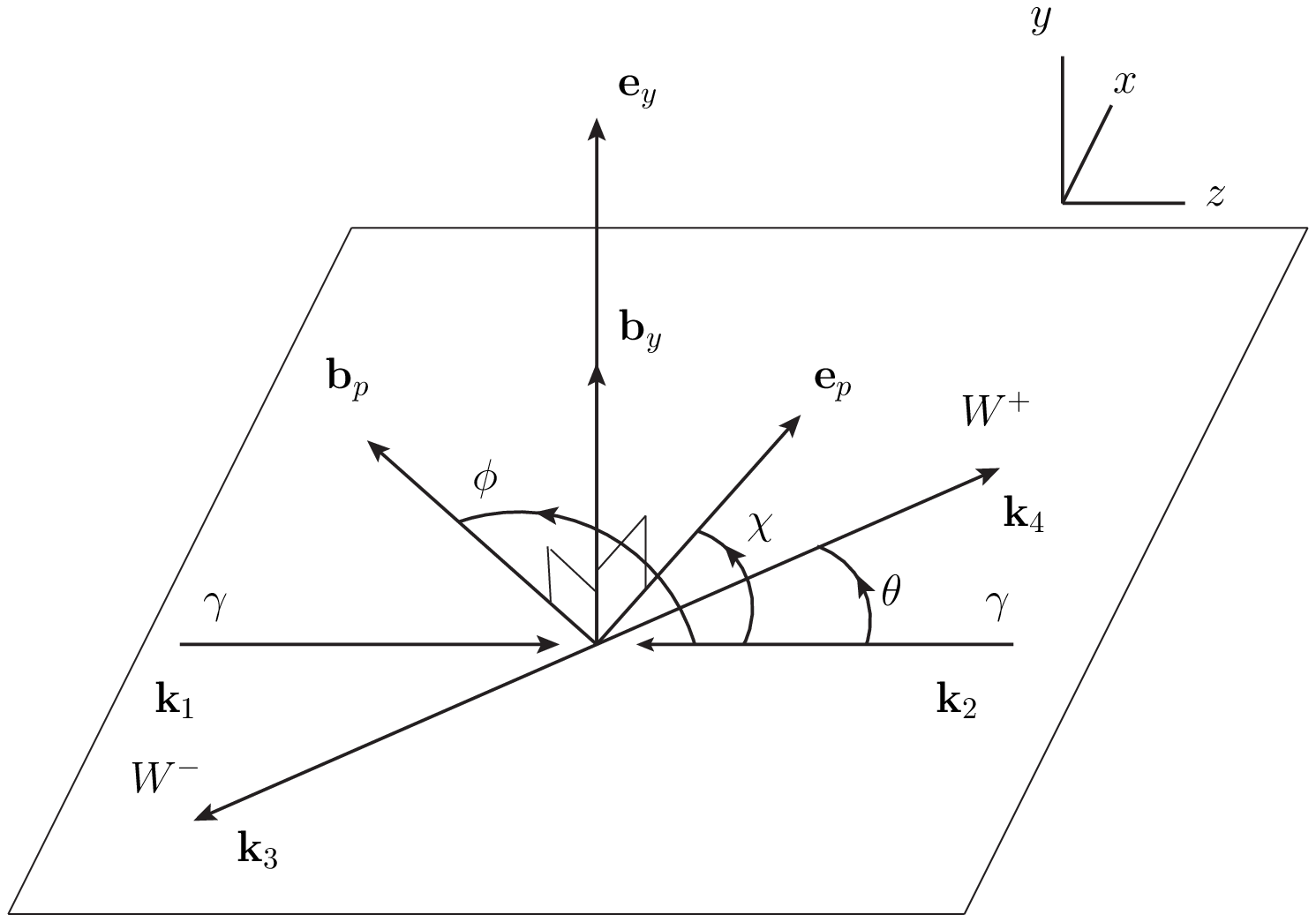}
\caption{\label{cinematica} The $\gamma\gamma\rightarrow WW$ reaction in the center of mass frame under the presence of a background field $b^{\alpha\beta}$. In this figure, $\textbf{e}_p$ and $\textbf{b}_p$ are the components of the $\textbf{e}$ and $\textbf{b}$ vectors, respectively, on the collision plane.}
\end{figure}
%%%%%%%%%%%%%%%%%%%%%%%%%%%%%%%%%%%%%%%%%%%%%%%%%%%%%%%%%%%%%%%%%%%%%%%%%%%%%%%%%%%%%%%%%%%%%%
In the framework of the SME, the pure anomalous contributions to the helicity amplitudes are
\begin{align}
{\cal M}^{b}_{\lambda_1\lambda_2\lambda_3\lambda_4} &= i e^{2}{\cal M}^{b}_{\mu\nu\lambda\rho} \epsilon^{\mu}(k_{1},\lambda_{1}) \epsilon^{\nu}(k_{2},\lambda_{2}) \epsilon^{\lambda *}(k_{3},\lambda_{3}) \epsilon^{\rho *}(k_{4},\lambda_{4}),\\
{\cal M}^{b}_{\lambda_1\lambda_2\lambda_3\lambda_{4}^{0}} &= i e^{2}{\cal M}^{b}_{\mu\nu\lambda\rho} \epsilon^{\mu}(k_{1},\lambda_{1}) \epsilon^{\nu}(k_{2},\lambda_{2}) \epsilon^{\lambda *}(k_{3},\lambda_{3}) \epsilon^{\rho *}(k_{4},\lambda_{4}^{0}),\\
{\cal M}^{b}_{\lambda_1\lambda_2\lambda_{3}^{0}\lambda_4} &= i e^{2}{\cal M}^{b}_{\mu\nu\lambda\rho} \epsilon^{\mu}(k_{1},\lambda_{1}) \epsilon^{\nu}(k_{2},\lambda_{2}) \epsilon^{\lambda *}(k_{3},\lambda_{3}^{0}) \epsilon^{\rho *}(k_{4},\lambda_{4}),\\
{\cal M}^{b}_{\lambda_1\lambda_2\lambda_{3}^{0}\lambda_{4}^{0}} &= i e^{2}{\cal M}^{b}_{\mu\nu\lambda\rho} \epsilon^{\mu}(k_{1},\lambda_{1}) \epsilon^{\nu}(k_{2},\lambda_{2}) \epsilon^{\lambda *}(k_{3},\lambda_{3}^{0}) \epsilon^{\rho *}(k_{4},\lambda_{4}^{0}).
\end{align}
The geometrical features of the collision are presented in Fig.~\ref{cinematica}. In this figure, the electric-like, $e^{i}\equiv\Lambda^{2}_{LV}b^{0i}$, and the magnetic-like, $b^{i}\equiv(1/2)\Lambda^{2}_{LV}\epsilon^{ijk}b^{jk}$, constant fields, have been decomposed into $\textbf{e}_p$, $\textbf{b}_p$ parallel components, and  $\textbf{e}_y$, $\textbf{b}_y$ perpendicular components, to the $x - z$ collision plane. Here, $\phi$ and $\chi$ are the angles formed by $\textbf{e}_p$ and $\textbf{b}_p$ with the $+z$ axis, respectively. In addition, we will make use of the following identity
\begin{equation}\label{abc}
a_{\alpha}\, b^{\alpha\beta}
\,c_{\beta}=\frac{1}{\Lambda_{LV}^{2}}\,[c_{0}\,\textbf{e}\cdot \textbf{a}-a_{0}\,\textbf{e}\cdot
\textbf{c}+\textbf{b}\cdot
(\textbf{a}\times \textbf{c})]\, ,
\end{equation}
which is valid for two arbitrary four-vectors $a_\mu$ and $c_\mu$.

After tedious algebraic manipulations, the corresponding helicity amplitudes can be expressed as
\begin{align}
{\cal M}^{b}_{\lambda_1\lambda_2\lambda_3\lambda_4} =& \frac{e^{2}s [E_{\lambda_1\lambda_2\lambda_3\lambda_4}^{y}e_{y} + B_{\lambda_1\lambda_2\lambda_3\lambda_4}^{p}b_{p} + i(E_{\lambda_1\lambda_2\lambda_3\lambda_4}^{p}e_{p} + B_{\lambda_1\lambda_2\lambda_3\lambda_4}^{y}b_{y})]}{128\Lambda^{2}_{LV}(\beta^{2}\cos^{2}\theta-1)},\\
{\cal M}^{b}_{\lambda_1\lambda_2\lambda_3\lambda_{4}^{0}} =& \frac{e^{2}s^{3/2} [E_{\lambda_1\lambda_2\lambda_3\lambda_{4}^{0}}^{y}e_{y} + B_{\lambda_1\lambda_2\lambda_3\lambda_{4}^{0}}^{p}b_{p} + i(E_{\lambda_1\lambda_2\lambda_3\lambda_{4}^{0}}^{p}e_{p} + B_{\lambda_1\lambda_2\lambda_3\lambda_{4}^{0}}^{y}b_{y})]}{64\sqrt{2}m_{W}\Lambda^{2}_{LV}(\beta^{2}\cos(2\theta)+\beta^{2}-2)},\\
{\cal M}^{b}_{\lambda_1\lambda_2\lambda_{3}^{0}\lambda_4} =&  \frac{e^{2}s^{3/2} [E_{\lambda_1\lambda_2\lambda_{3}^{0}\lambda_4}^{y}e_{y} + B_{\lambda_1\lambda_2\lambda_{3}^{0}\lambda_4}^{p}b_{p} + i(E_{\lambda_1\lambda_2\lambda_{3}^{0}\lambda_4}^{p}e_{p} + B_{\lambda_1\lambda_2\lambda_{3}^{0}\lambda_4}^{y}b_{y})]}{64\sqrt{2}m_{W}\Lambda^{2}_{LV}(\beta^{2}\cos(2\theta)+\beta^{2}-2)},\\
{\cal M}^{b}_{\lambda_1\lambda_2\lambda_{3}^{0}\lambda_{4}^{0}} =&  \frac{e^{2}s^{2} [E_{\lambda_1\lambda_2\lambda_{3}^{0}\lambda_{4}^{0}}^{y}e_{y} + B_{\lambda_1\lambda_2\lambda_{3}^{0}\lambda_{4}^{0}}^{p}b_{p} + i(E_{\lambda_1\lambda_2\lambda_{3}^{0}\lambda_{4}^{0}}^{p}e_{p} + B_{\lambda_1\lambda_2\lambda_{3}^{0}\lambda_{4}^{0}}^{y}b_{y})]}{32 m_{W}^{2}\Lambda^{2}_{LV}(\beta^{2}\cos(2\theta)+\beta^{2}-2)}.
\end{align}
The expressions for $E^{p,y}$ and $B^{p,y}$ are presented in the Appendix~\ref{APB}.

It should be noted that all the SME helicity amplitudes are different from zero. This implies that we have polarization states in which the new physics effect that violates Lorentz symmetry appears free of the SM contribution. Notice that even though we have polarization states where there is no SM contribution, these same polarization states are different from zero in the CESM, so LV signal cannot be obtained cleanly from SM or CESM contribution.

\section{Discussion}\label{D}
This section presents numerical results for the $\gamma\gamma\to WW$ reaction. For simplicity reasons, the helicity amplitudes are organized in the same way they were presented in reference~\cite{RSM}, where only the sum of two transverse polarizations of the $W$ boson is considered. Accordingly, helicity states are indicated by four labels (from left to right), the first two correspond to the photons, and the remaining two refer to the $W$ bosons. The labels $-$,$+$ represent left-handed and right-handed photons, respectively, $L$ indicates longitudinal $W$ boson polarization, and $T$ symbolizes the sum of two transverse $W$ boson polarizations: for instance, ${\cal{M}}_{+,+,L,T}={\cal{M}}_{+,+,0,-}+{\cal{M}}_{+,+,0,+}$. We have computed all the polarization state contributions, but due to the new physics effects provide marginal contributions for the $(\pm,\mp,L,L)$, $(\pm,\pm,T,T)$, and $(\pm,\mp,T,T)$ polarization states, we will only analyze the behavior of differential cross section for the following polarization states: $(\pm,\pm,L,L)$, $(\pm,\mp,(L,T+T,L))$, $(\pm,\pm,(L,T+T,L))$. For simplicity, our study is reduced to consider new physics effects at the tree level, so that SM one-loop level corrections are not taken into account.

The new physics effects that disregard the Lorentz symmetry offer additional information which could be even more important than the new physics that respects this symmetry. The former can be evidenced not only by the relative value of the new physics scale but also due to privileged directions determined by the $\textbf{e}, \textbf{b}$ constant background fields. In order to find possible scenarios of Lorentz symmetry violation we will study in detail the differential cross section for the $\gamma \gamma \to WW$ process. It is important to emphasize that simultaneous information about dependence of privileged angular directions of the background fields and scattering angle in the total cross section is lost, since the scattering angle has been already integrated. Thus, we present a close examination of the differential cross section. One of the questions we want to answer is whether LV is more sensitive to the $\textbf{b}$ or to the $\textbf{e}$ background fields. We are interested in the search for scenarios in which either the SM contribution is absent or the new physics effects differ significantly from it. In this regard, we look for optimal values for the Lorentz violating parameters which enhance the new physics effects arising from the SME, assuming that $|\textbf{e}|\sim 1$ and $|\textbf{b}|\sim1$. Therefore, we will analyze the following scenarios: a) $\textbf{e}=0$, $\textbf{b}\neq 0$, b) $\textbf{e}\neq0$, $\textbf{b}=0$, and c) $\textbf{e}\neq0$, $\textbf{b}\neq 0$.

As already mentioned above, we are interested in contrasting new physics effects arising in the CESM approach or in SME one, since they could be observed at ILC. To make predictions, some values for the parameters of the CESM, $(\Lambda, \alpha_W)$, and for the ones of the SME, $(\Lambda_{LV}, e_p, e_y, b_p, b_y, \chi, \phi)$, must be assumed. In a previous work carried out by some of us, a constraint given by $\Lambda_{LV}>1.96$ TeV on the Lorentz-violating scale associated with the ${\cal O}^W$ operator, was obtained from experimental data on the $B\to X_s\gamma$ decay~\cite{T1}. For comparison purposes, in the following we will assume that $\Lambda=\Lambda_{LV}=2$ TeV, besides, we will assume that $\alpha_W=1$.

\subsection{Differential cross section}
We will discuss our analysis of the $\gamma \gamma \to WW$ differential cross section on the scattering angle interval $20^\circ<\theta<160^\circ$, which is usually employed in the experimental setting and it will give us relevant information for search of new physics, since it will indicate what are the most promising angular regions to look for Lorentz violation.

\subsubsection{Scenario $\textbf{e}=0$, $\textbf{b}\neq 0$}
For the three polarization states above mentioned, we will use the values $b_y = b_p =1$ for the $\textbf{b}$ constant background field.\\
%%%%%%%%%%%%%%%%%%%%%%%%%%%%%%%%%%%%%%%%%%%%%%%%%%%%%%%%%%%%%%%%%%%%%%%%%%%
\begin{figure}[htb!]
\centering
\subfigure[]{\includegraphics[scale=0.65]{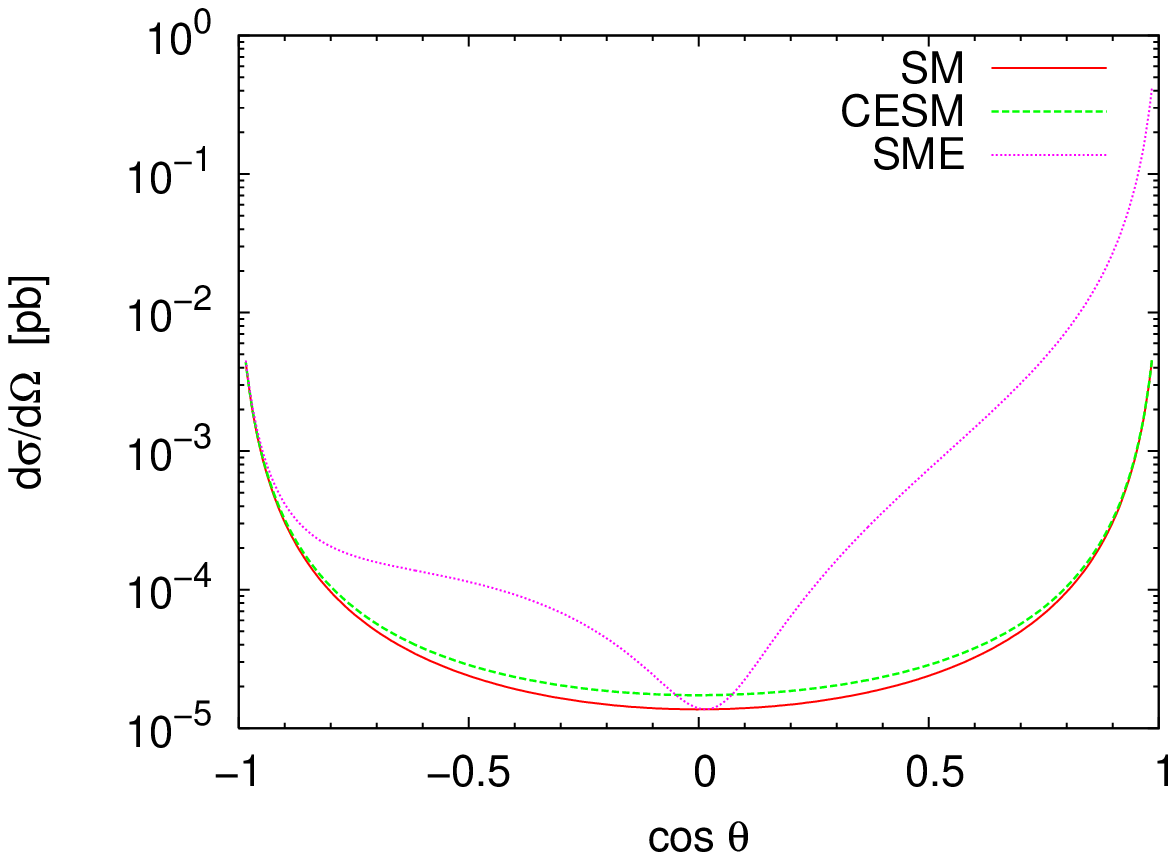}}\qquad
\subfigure[]{\includegraphics[scale=0.65]{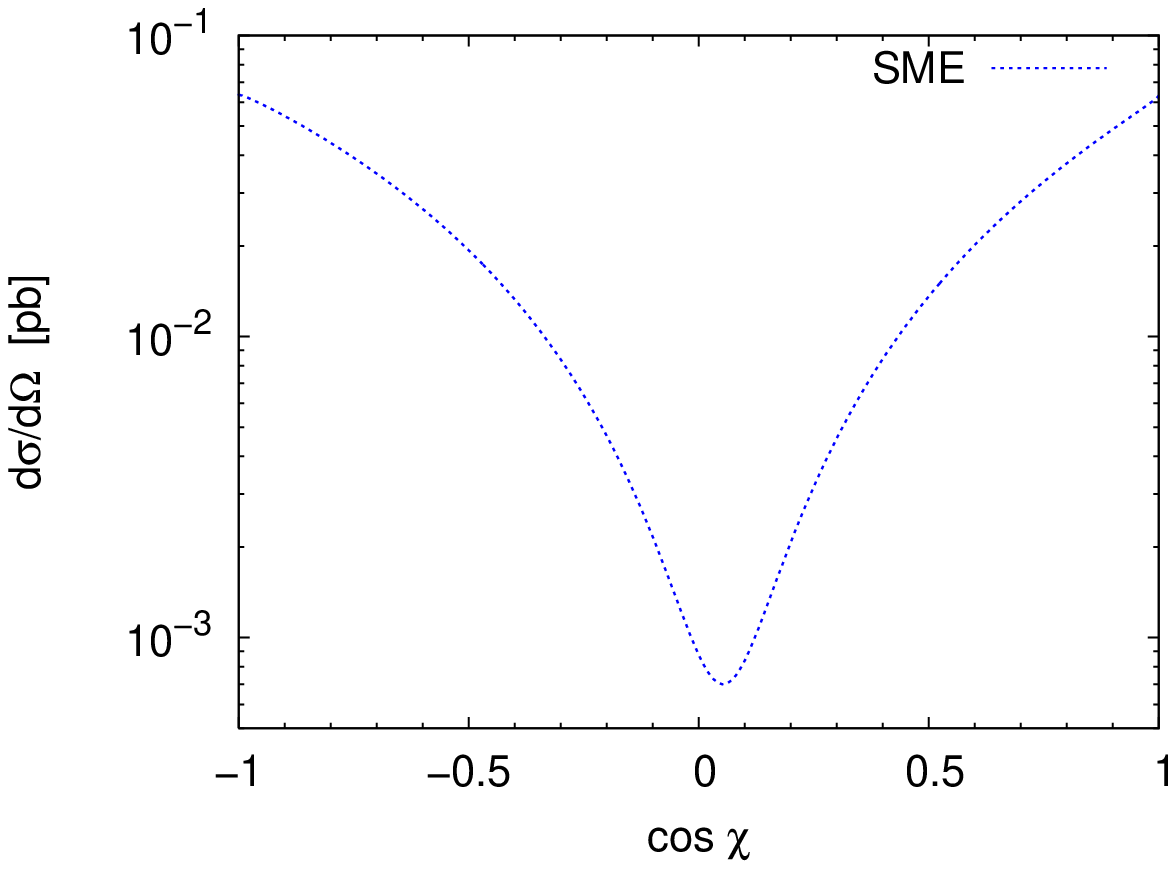}}
\caption{Differential cross section for the $\gamma\gamma\rightarrow WW$ process with the $(\pm,\pm ,L,L)$ polarization state at $\sqrt{s}=1$ TeV ($\textbf{e}=0$, $\textbf{b}\neq 0$). (a) $\chi=176_{\cdot}96^\circ$. (b) $\theta=20^\circ$.}
\label{ppLLb}
\end{figure}
\textbf{The $(\pm,\pm,L,L)$ polarization}.
In this type of polarization state new physics effects that come from the SME become up to 2 orders of magnitude larger than the signal of the SM. Notice that the deviation of the CESM regarding the SM contribution is virtually negligible. Figure~\ref{ppLLb} presents the polarized differential cross section as a function of the scattering angle $\theta$ and the $\chi$ angle. In Fig.~\ref{ppLLb}(a), it can clearly be seen that the differential cross section reaches its maximum value at $\chi=176_{\cdot}96^\circ$ for $\theta=20^\circ$, which is $5\times10^{-2}$ pb. In Fig.~\ref{ppLLb}(b) it can be observed that the differential cross section reaches its maximum value near the ends of the $\chi$ angular interval, specifically, the maximum value corresponds to $\chi=176_{\cdot}96^\circ$. From the previous analysis it can be concluded that LV signal is more intense in the vicinity of $\theta=20^\circ$ and $\chi=176_{\cdot}96^\circ$.

\begin{figure}[htb!]
\centering
\subfigure[]{\includegraphics[scale=0.65]{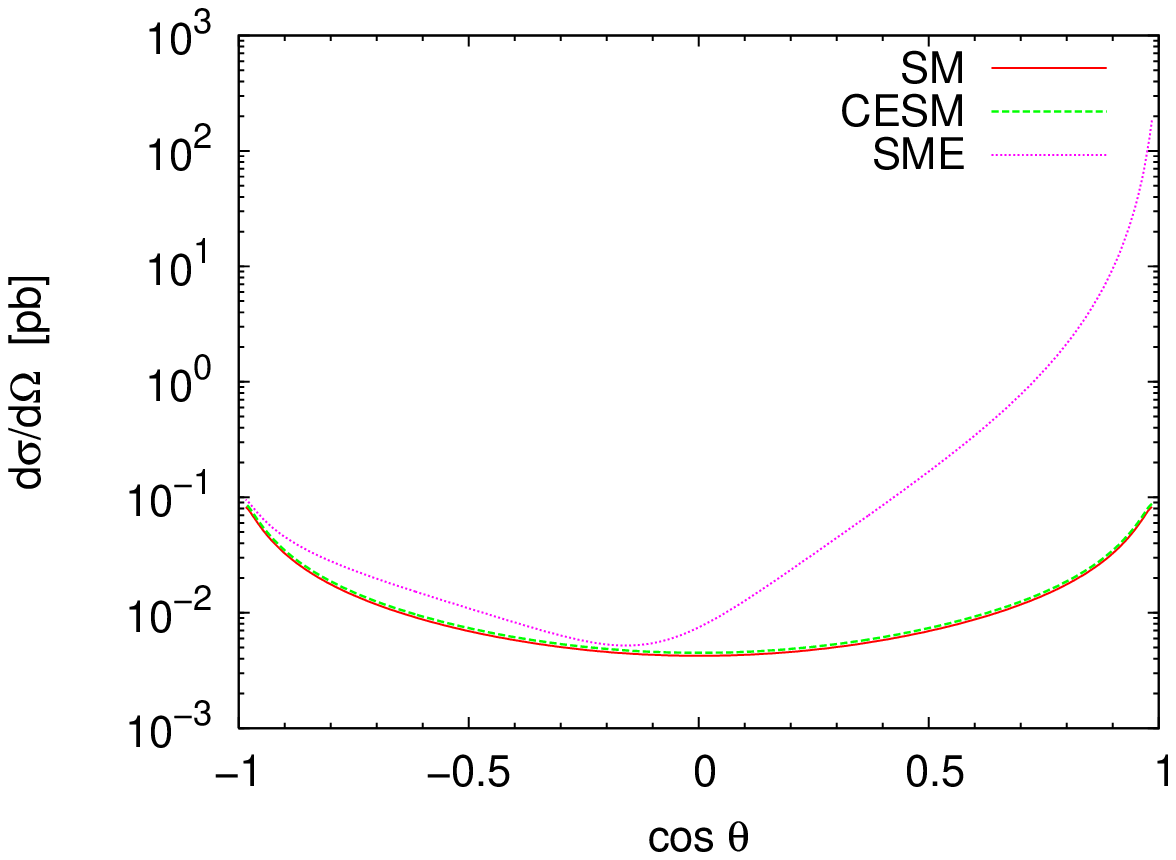}}\qquad
\subfigure[]{\includegraphics[scale=0.65]{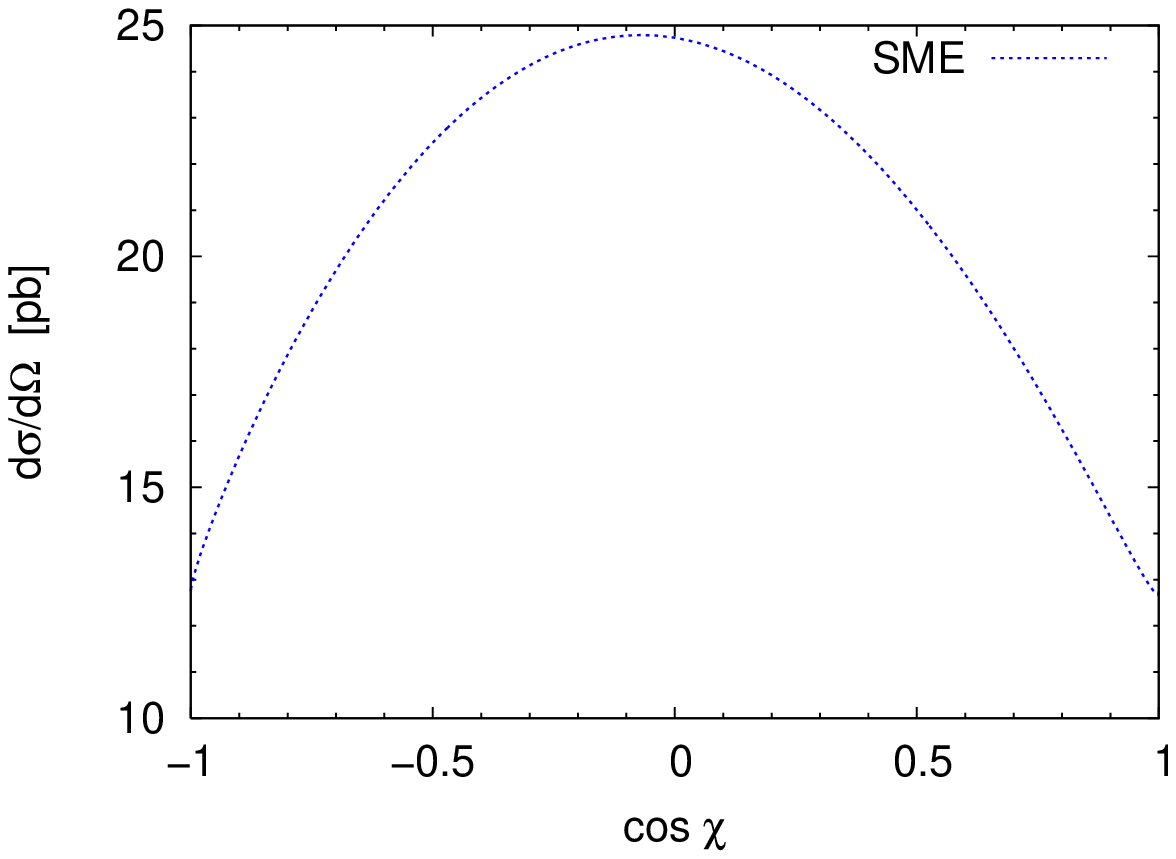}}
\caption{Differential cross section for the $\gamma\gamma\rightarrow WW$ process with the $(\pm,\mp, (L,T+T,L))$ polarization state at $\sqrt{s}=1$ TeV ($\textbf{e}=0$, $\textbf{b}\neq 0$). (a) $\chi=97_{\cdot}41^{\circ}$. (b) $\theta=20^{\circ}$.}
\label{pnLTTLb}
\end{figure}

\textbf{The $(\pm,\mp,(L,T+T,L))$ polarization}.
Here, the contribution of the SME can be up to 3 orders of magnitude larger than the respective signal of the SM. In Fig.~\ref{pnLTTLb}, we show the polarized differential cross section as a function of the scattering angle $\theta$ and the $\chi$ angle.
It should be noted that there is no appreciable deviation between the SM contribution when compared with the CESM contribution. From Fig.~\ref{pnLTTLb}(a), we can observe that the maximum value of the differential cross section is located around $\theta=20^\circ$ for $\chi=97_{\cdot}41^{\circ}$, being 24.76 pb. In Fig.~\ref{pnLTTLb}(b), it can be noted that the differential cross section is maximal when $\chi=97_{\cdot}41^{\circ}$. Thus, new physics effects coming from the SME are favored for angular regions in the neighborhood of $\theta=20^\circ$ and $\chi=97_{\cdot}41^{\circ}$.
%%%%%%%%%%%%%%%%%%%%%%%%%%%%%%%%%%%%%%%%%%%%%%%%%%%%%%%%%%%%%%%%%%%%%%%%%%%%%%%%%%%%%%%%%%%%%%
%%%%%%%%%%%%%%%%%%%%%%%%%%%%%%%%%%%%%%%%%%%%%%%%%%%%%%%%%%%%%%%%%%%%%%%%%%%%%%%%%%%%%%%%%%%%%%
%%%%%%%%%%%%%%%%%%%%%%%%%%%%%%%%%%%%%%%%%%%%%%%%%%%%%%%%%%%%%%%%%%%%%%%%%%%%%%%%%%%%%%%%%%%%%%
\begin{figure}[htb!]
\centering
\subfigure[]{\includegraphics[scale=0.65]{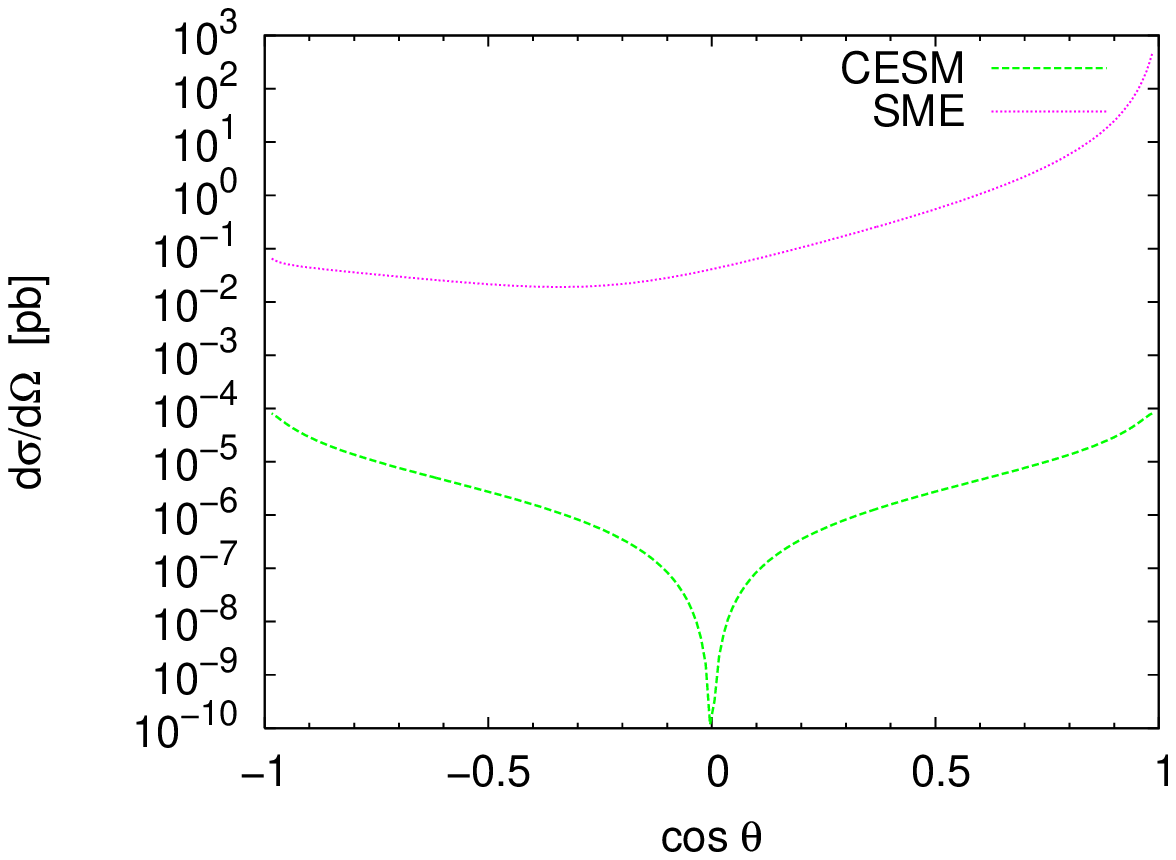}}\qquad
\subfigure[]{\includegraphics[scale=0.65]{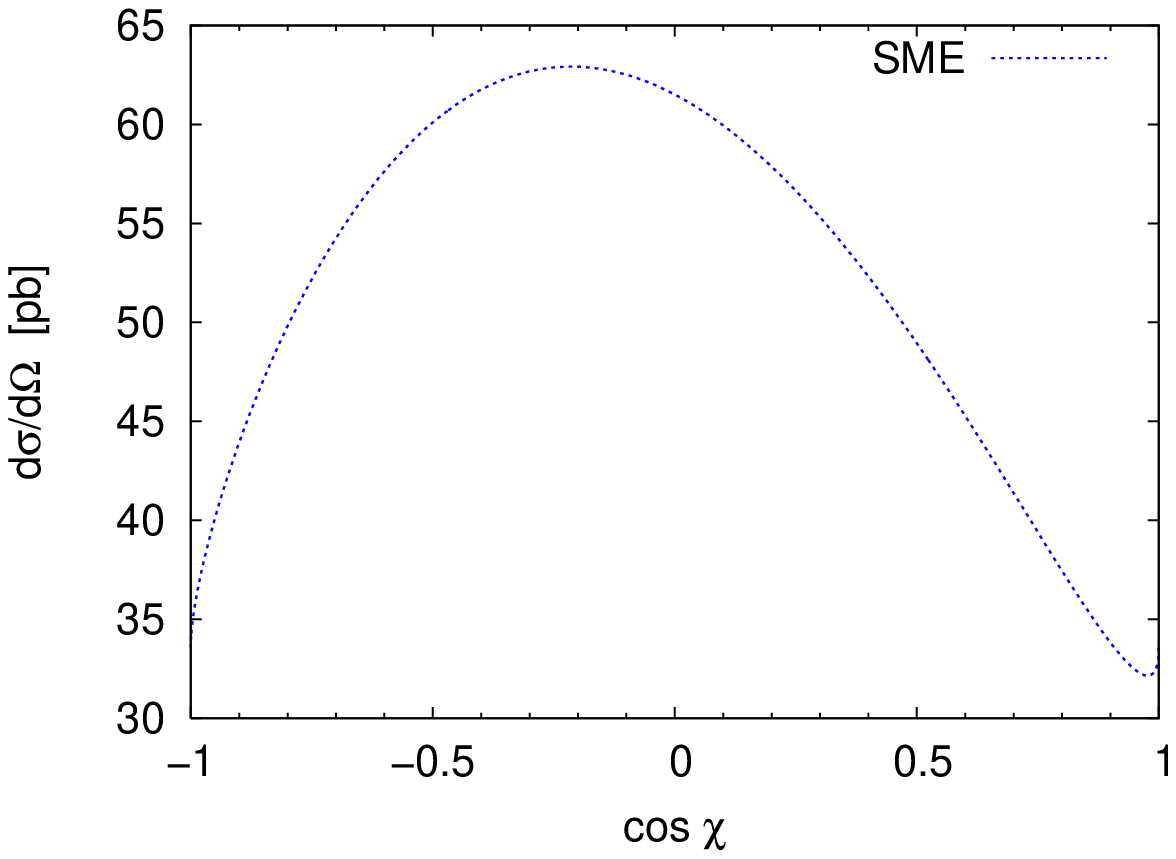}}
\caption{Differential cross section for the $\gamma\gamma\rightarrow WW$ process with the $(\pm,\pm, (L,T+T,L))$ polarization state at $\sqrt{s}=1$ TeV ($\textbf{e}=0$, $\textbf{b}\neq 0$). (a) $\chi=102_{\cdot}36^{\circ}$. (b) $\theta=20^{\circ}$.}
\label{ppLTTLb}
\end{figure}

\textbf{The $(\pm,\pm,(L,T+T,L))$ polarization}. This polarization state is very interesting because there are only present new physics effects, since the tree-level SM contribution is exactly zero. Therefore, this particular polarization state results in a good framework to confront the new physics effects coming from CESM and SME. In Fig.~\ref{ppLTTLb}, the $(\pm,\pm,(L,T+T,L))$ differential cross section is exhibited as a function of the scattering angle $\theta$ and the $\chi$ angle. In Fig.~\ref{ppLTTLb}(a), it is clearly appreciable that the differential cross section reaches its maximum value at $\theta=20^\circ$ for $\chi=102_{\cdot}36^{\circ}$, being of the order of $10^{2}$ pb. Based on Fig.~\ref{ppLTTLb}(b), it can be discerned that new physics effects arising from SME are magnified in the proximity of $\theta=20^\circ$ and $\chi=102_{\cdot}36^{\circ}$. This case is relevant because the detection of new physics effects with Lorentz symmetry violation can be observed more easily, since as mentioned above, the contribution of SM is zero to this order of perturbation theory, in addition, the CESM signal is very suppressed.

\subsubsection{Scenario $\textbf{e}\neq0$, $\textbf{b}=0$}
In this scenario we will use $e_p=e_y=1$ for the parallel and normal components of the $\textbf{e}$ constant background field.

%%%%%%%%%%%%%%%%%%%%%%%%%%%%%%%%%%%%%%%%%%%%%%%%%%%%%%%%%%%%%%%%%%%%%%%%%%%%%%%%%%%%%%%%%%%%%%
\begin{figure}[htb!]
\centering
\subfigure[]{\includegraphics[scale=0.65]{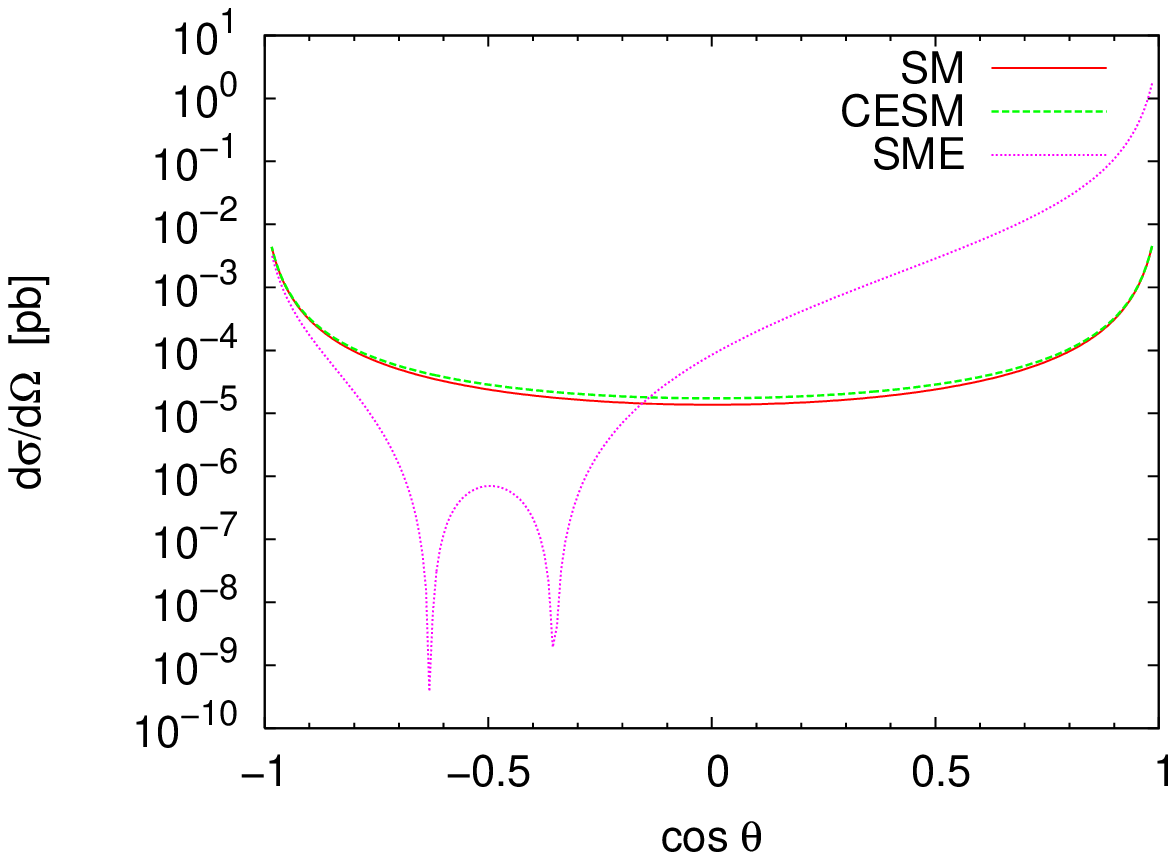}}\qquad
\subfigure[]{\includegraphics[scale=0.65]{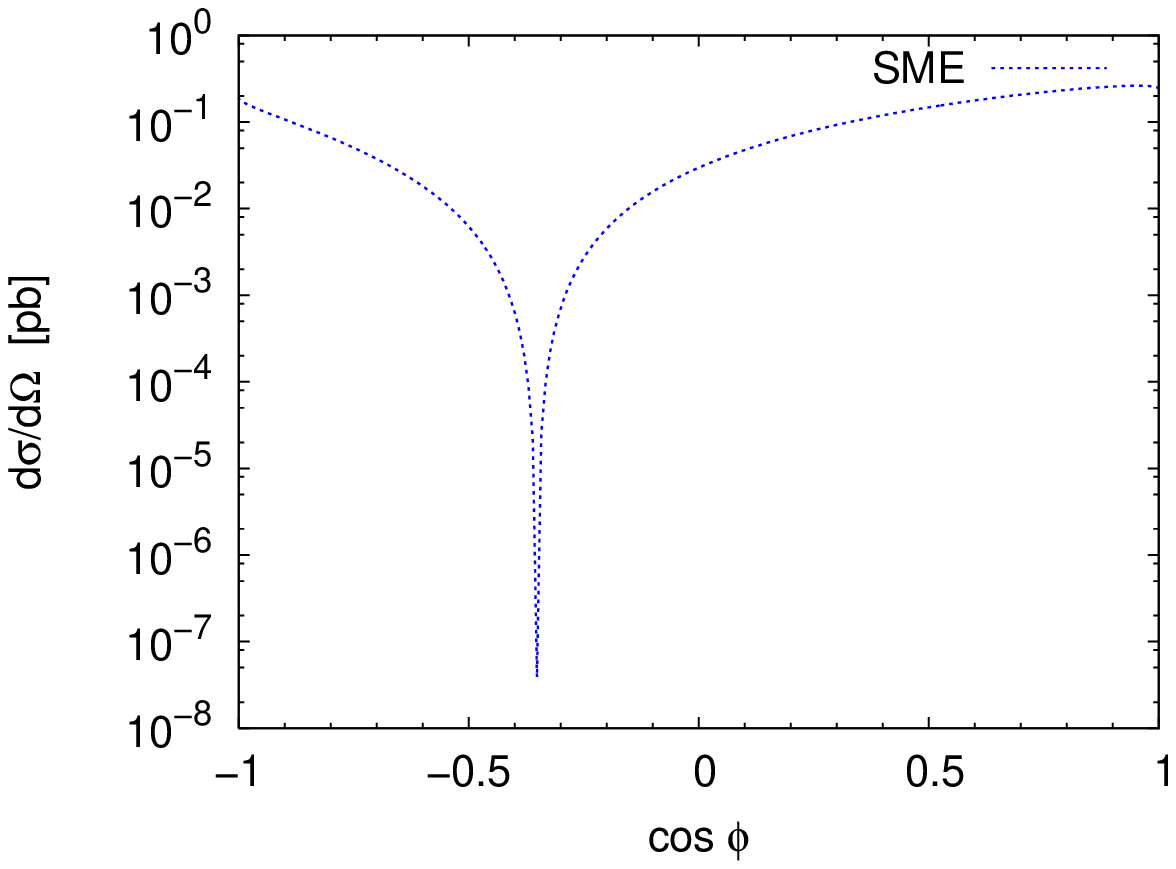}}
\caption{Differential cross section for the $\gamma\gamma\rightarrow WW$ process with the $(\pm,\pm, L,L)$ polarization state at $\sqrt{s}=1$ TeV ($\textbf{e}\neq 0$, $\textbf{b}=0$). (a) $\phi=17_{\cdot}48^{\circ}$. (b) $\theta=20^{\circ}$.}
\label{ppLLe}
\end{figure}
%%%%%%%%%%%%%%%%%%%%%%%%%%%%%%%%%%%%%%%%%%%%%%%%%%%%%%%%%%%%%%%%%%%%%%%%%%%%%%%%%%%%%%%%%%%%%%

\textbf{The $(\pm,\pm,L,L)$ polarization}. In this case, it must be stressed that there is no appreciable difference between SM and CESM contributions. However, Lorentz-violating effects arising from the SME in certain angular regions are about 2 orders of magnitude larger than the respective SM signal. In Fig.~\ref{ppLLe}, the $(\pm,\pm,L,L)$ polarized differential cross section is shown as a function of the scattering angle $\theta$ and the $\phi$ angle. From this figure, it can be observed that the differential cross section is of the order of $10^{-1}$ pb in the vicinity of $\phi=17_{\cdot}48^{\circ}$ and $\theta=20^\circ$. As it can be inferred from these figures, Lorentz-violating effects are dominant for this kind of polarization, moreover, its maximum intensity does not exceed one order of magnitude as compared to SME maximum in the previous case ($5\times10^{-2}$ pb when $\textbf{e}= 0$, $\textbf{b}\neq 0$), for the same polarization state.

%%%%%%%%%%%%%%%%%%%%%%%%%%%%%%%%%%%%%%%%%%%%%%%%%%%%%%%%%%%%%%%%%%%%%%%%%%%%%%%%%%%%%%%%%%%%%%
\begin{figure}[htb!]
\centering
\subfigure[]{\includegraphics[scale=0.65]{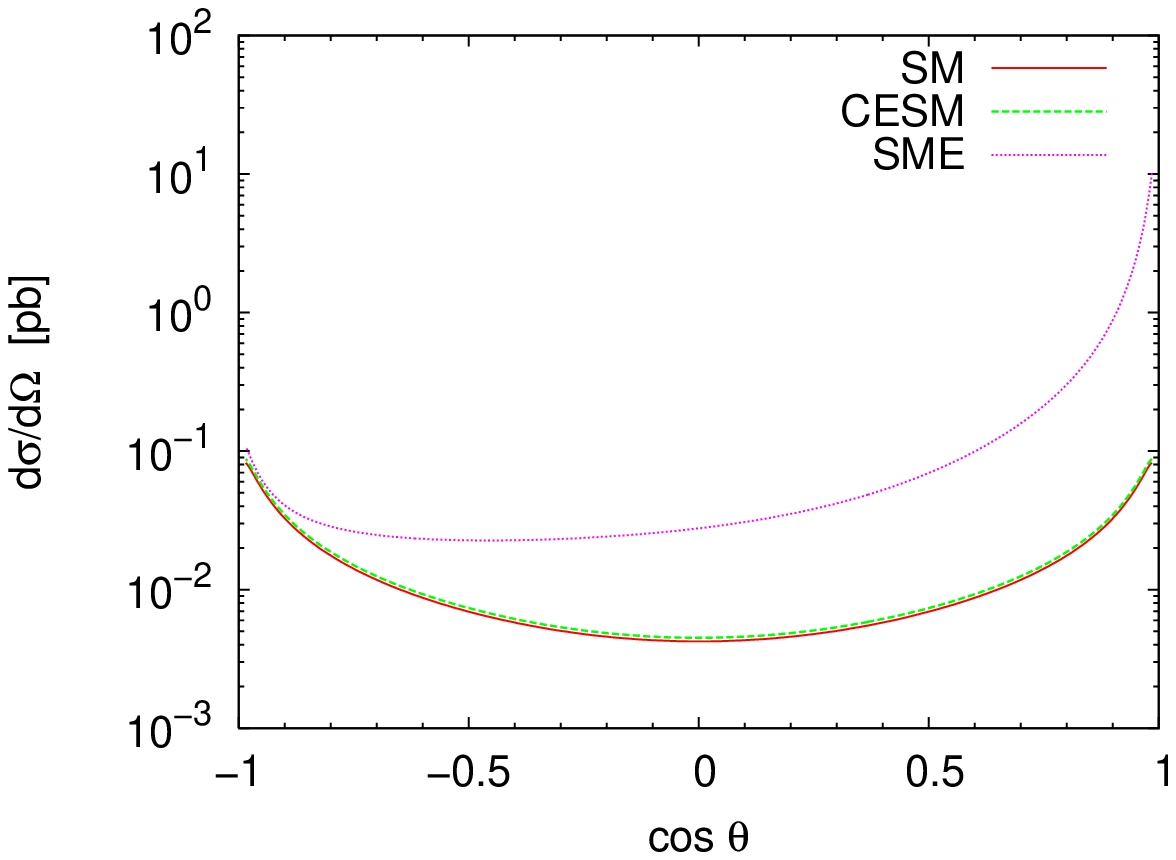}}\qquad
\subfigure[]{\includegraphics[scale=0.65]{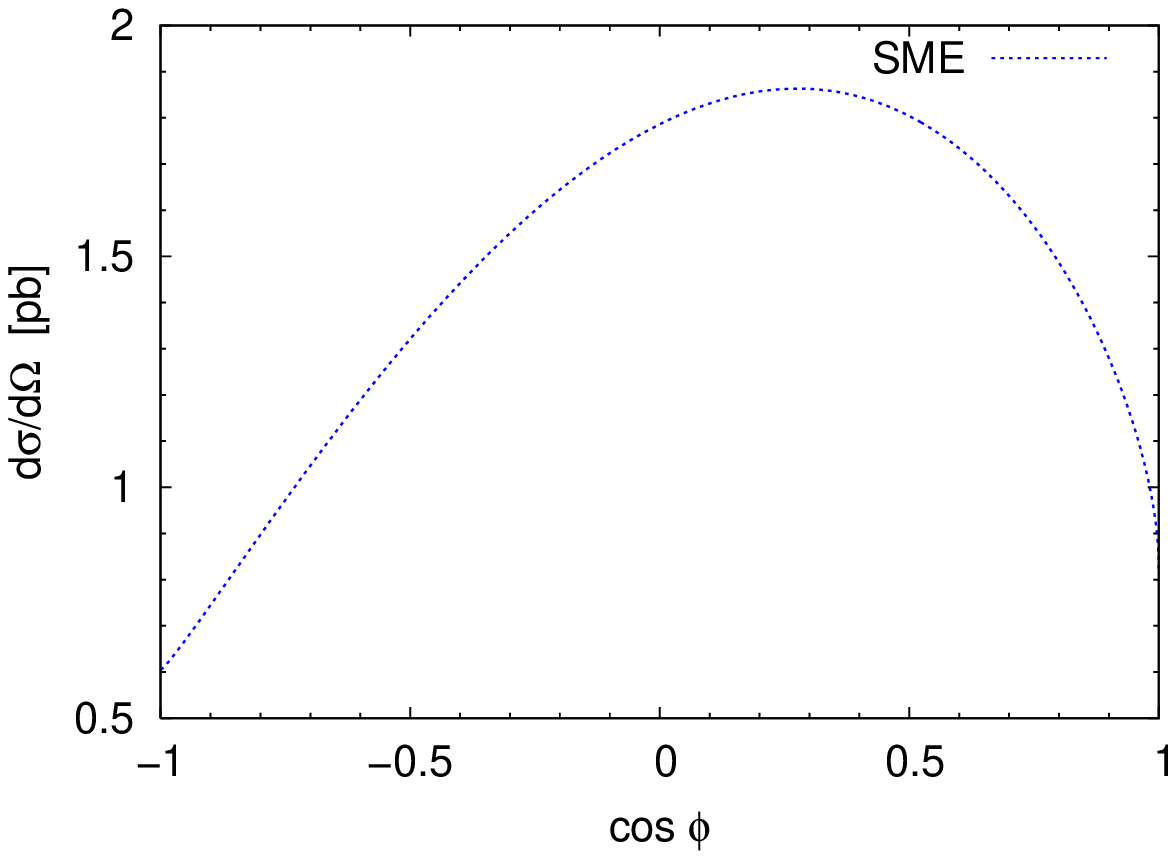}}
\caption{Differential cross section for the $\gamma\gamma\rightarrow WW$ process with the $(\pm,\mp, (L,T+T,L))$ polarization state at $\sqrt{s}=1$ TeV ($\textbf{b}=0$, $\textbf{e}\neq 0$). (a) $\phi=73_{\cdot}99^{\circ}$. (b) $\theta=20^{\circ}$.}
\label{pnLTTLe}
\end{figure}
%%%%%%%%%%%%%%%%%%%%%%%%%%%%%%%%%%%%%%%%%%%%%%%%%%%%%%%%%%%%%%%%%%%%%%%%%%%%%%%%%%%%%%%%%%%%%%

\textbf{The $(\pm,\mp,(L,T+T,L))$ polarization}. Fig.~\ref{pnLTTLe} presents the $(\pm,\mp,(L,T+T,L))$ polarized differential cross section as a function of the scattering angle $\theta$ and the $\phi$ angle. From this figure, it can be appreciated that in the best situation the SME contribution is above the SM signal in around 2 orders of magnitude. It should be noted that there are no tangible differences between the contributions of the SM and the CESM. In Fig.~\ref{pnLTTLe}, we can see that the differential cross sections reaches at most $2$ pb in the neighborhood of $\phi=162_{\cdot}19^{\circ}$ for $\theta=20^\circ$; indeed, the precise value of the maximum corresponds to 1.86 pb. Therefore, to this polarization state, the best place to search for LV corresponds to the $\textbf{e}= 0$, $\textbf{b}\neq 0$ case, because the largest value for its differential cross section is one order of magnitude larger than the maximum differential cross section obtained in the $\textbf{e}\neq 0$, $\textbf{b}=0$ scenario.

%%%%%%%%%%%%%%%%%%%%%%%%%%%%%%%%%%%%%%%%%%%%%%%%%%%%%%%%%%%%%%%%%%%%%%%%%%%%%%%%%%%%%%%%%%%
\begin{figure}[htb!]
\centering
\subfigure[]{\includegraphics[scale=0.65]{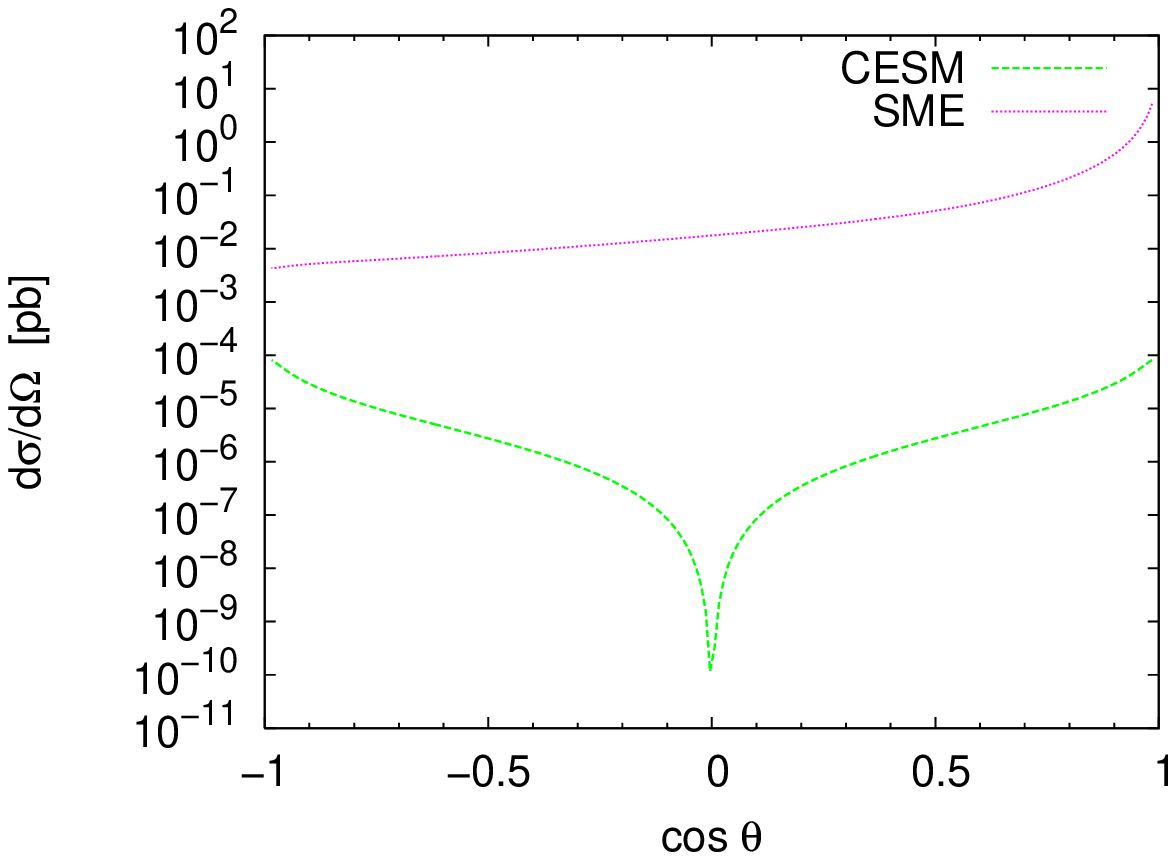}}\qquad
\subfigure[]{\includegraphics[scale=0.65]{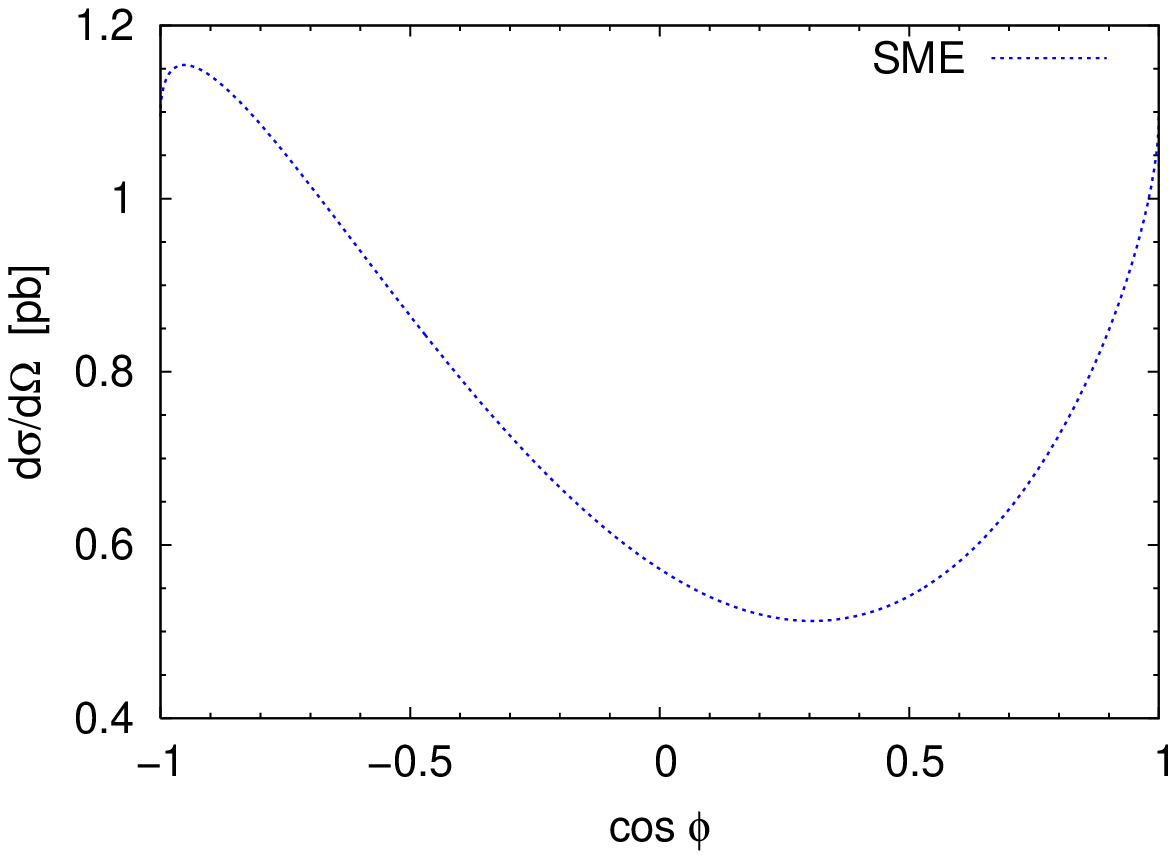}}
\caption{Differential cross section for the $\gamma\gamma\rightarrow WW$ process with the $(\pm,\pm, (L,T+T,L))$ polarization state at $\sqrt{s}=1$ TeV ($\textbf{b}=0$, $\textbf{e}\neq 0$). (a) $\phi=162.19^{\circ}$. (b) $\theta=20^{\circ}$.}
\label{ppLTTLe}
\end{figure}
%%%%%%%%%%%%%%%%%%%%%%%%%%%%%%%%%%%%%%%%%%%%%%%%%%%%%%%%%%%%%%%%%%%%%%%%%%%%%%%%%%%%%%%%%%%

\textbf{The $(\pm,\pm,(L,T+T,L))$ polarization}. It should be recalled that there is no contributions from SM to this polarization state. In Fig.~\ref{ppLTTLe} we display the $(\pm,\pm,(L,T+T,L))$ polarized differential cross section as a function of the scattering angle and the $\phi$ angle. From the Figs.~\ref{ppLTTLe}(a) and  \ref{ppLTTLe}(b), it can be seen a very clear signal of LV in the proximity of $\theta=20^\circ$ and $\phi=73_{\cdot}99^{\circ}$, which has an intensity of 10 pb. Even though this polarization state is suitable to study possible LV signals, it is important to point out that the maximum in differential cross section is one order of magnitude below the same polarization state result when $\textbf{e}= 0$, $\textbf{b}\neq 0$.

In summary, we can say in general terms that the Lorentz-violating effects are disadvantaged in the context of this scenario if they are compared with the previous scenario. Accordingly, as in the study of Lorentz-violating effects on the $e\gamma\to W\nu_e$ process~\cite{Tla}, it becomes manifest the dominant effect of the $\mathbf{b}$ background field on the impact of Lorentz symmetry violation.

%it is evidenced that the effects of Lorentz symmetry violation are more sensitive to the presence of the $\mathbf{b}$ constant background field.

%%%%%%%%%%%%%%%%%%%%%%%%%%%%%%%%%%%%%%%%%%%%%%%%%%%%%%%%%%%%%%%%%%%%%%%%%%%%%%%%%%%%%%%

\subsubsection{Scenario $\textbf{e}\neq0$, $\textbf{b}\neq0$}
One of the main objectives of this work consists in finding angular localities for the differential cross sections where signals of LV can be isolated not only from the SM contribution but also from other sources of anomalous effects. In the previous two scenarios analyzed, we have found that signals of LV can be clearly observed in the scenario $\textbf{e}=0$, $\textbf{b}\neq0$, however, although such effects are also seen in the $\textbf{e}\neq 0$, $\textbf{b}=0$ case, they are much less intense. Concretely, the values of the differential cross sections in the former scenario are, in general terms, one order of magnitude larger than in the latter one. So, one should expect that a more general scenario with both non-vanishing electric-like and magnetic-like constant background fields does not modify essentially the prediction of the dominant scenario $\textbf{e}=0$, $\textbf{b}\neq0$. To foresee any subtle cancellation coming from interference effects between $\mathbf{e}$ and $\mathbf{b}$, we have performed an exhaustive analysis showing that predictions in the $\textbf{e}=0$, $\textbf{b}\neq0$ scenario remain essentially unchanged.

% % % % % % % % % % % % % % % % % % % % % % % % % % % % % % % % % % %
\subsection{Total cross section}
In this part, we will focus our attention on angular regions where Lorentz symmetry violation effects stand out notoriously. Thus, we will only study the $\textbf{e}=0$, $\textbf{b}\neq0$ scenario since it is the most promising one. As it was demonstrated in previous sections, signals of Lorentz violation are magnified in the vicinity of $\theta=20^\circ$ and $\chi\simeq100^\circ$ for differential cross sections with polarization states $(\pm,\mp,(L,T+T,L))$ and $(\pm,\pm,(L,T+T,L))$. In the case of differential cross section $(\pm,\pm,L,L)$, it should be remembered that a transparent LV signal appears in the proximity of $\chi=176_{\cdot}96^\circ$ and $\theta=20^\circ$, but, in general, it is suppressed in at least one order of magnitude with respect to the $(\pm,\mp,(L,T+T,L))$ and $(\pm,\pm,(L,T+T,L))$ ones. Thereby, to better appreciate the Lorentz-violating effects, we will integrate the differential cross sections $(\pm,\mp,(L,T+T,L))$ and $(\pm,\pm,(L,T+T,L))$ in the angular interval $20^\circ$ $<\theta<$ $40^\circ$, for $\chi=100^\circ$. In the case of the polarized cross section $(\pm,\pm,L,L)$, we will consider the region $20^\circ$ $<\theta<$ $40^\circ$, with $\chi=176_{\cdot}96^\circ$.
%%%%%%%%%%%%%%%%%%%%%%%%%%%%%%%%%%%%%%%%%%%%%%%%%%%%%%%%%%%%%%%%%%%%%%%%%%%%%%%%%%%%%%%%%%%%%%
\begin{figure}[htb!]
\centering
\includegraphics[scale=0.7]{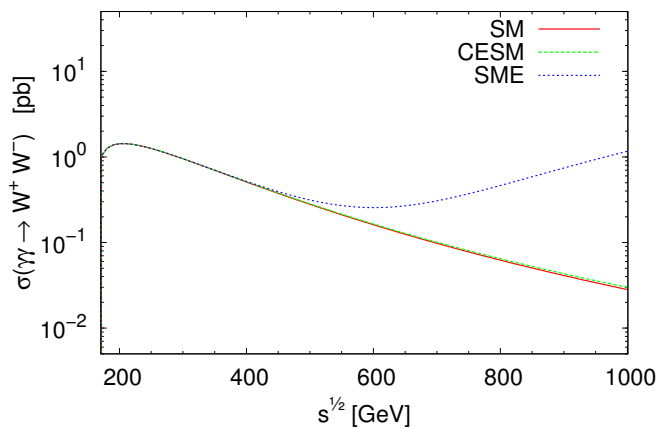}
\caption{Polarized cross section $(\pm,\mp, (L,T+T,L))$ for the $\gamma\gamma\to WW$ process; integrated in the angular interval $20^\circ<\theta< 40^\circ$. Only the dominant scenario $\textbf{e}=0$, $\textbf{b}\neq 0$, with $\chi=100^\circ$, is considered.}
\label{SEpnLTTL}
\end{figure}
%%%%%%%%%%%%%%%%%%%%%%%%%%%%%%%%%%%%%%%%%%%%%%%%%%%%%%%%%%%%%%%%%%%%%%%%%%%%%%%%%%%%%%%%%%%%%%

In Fig.~\ref{SEpnLTTL}, the behavior of the integrated cross section $(\pm,\mp, (L,T+T,L))$ in the above cited angular region is displayed as a function of the center of mass energy. From this figure, a clear signal of LV can be observed starting in $\sqrt{s}\simeq500$ GeV, which can reaches values close to 2 orders of magnitude over the SM (CESM) signal for $\sqrt{s}\simeq1000$ GeV.
%%%%%%%%%%%%%%%%%%%%%%%%%%%%%%%%%%%%%%%%%%%%%%%%%%%%%%%%%%%%%%%%%%%%%%%%%%%%%%%%%%%%%%%%%%%%%%
\begin{figure}[htb!]
\centering
\includegraphics[scale=0.7]{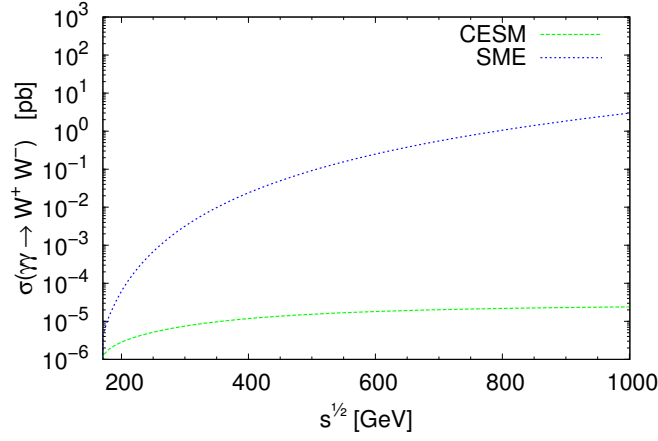}
\caption{Polarized cross section $(\pm,\pm, (L,T+T,L))$ for the $\gamma\gamma\to WW$ process; integrated in the angular interval $20^\circ<\theta< 40^\circ$. Only the dominant scenario $\textbf{e}=0$, $\textbf{b}\neq 0$, with $\chi=100^\circ$, is considered.}
\label{SEppLTTL}
\end{figure}
%%%%%%%%%%%%%%%%%%%%%%%%%%%%%%%%%%%%%%%%%%%%%%%%%%%%%%%%%%%%%%%%%%%%%%%%%%%%%%%%%%%%%%%%%%%%%%

Figure~\ref{SEppLTTL} presents the behavior of the polarized cross section $(\pm,\pm, (L,T+T,L))$ in the energy region 200 GeV$<\sqrt{s}<$1200 GeV. In these plots, it can be seen that the SME contribution dominates by, at least, 5 orders of magnitude over the CESM signal for $\sqrt{s}=1000$ GeV (there is no SM contribution).

%%%%%%%%%%%%%%%%%%%%%%%%%%%%%%%%%%%%%%%%%%%%%%%%%%%%%%%%%%%%%%%%%%%%%%%%%%%%%%%%%%%%%%%%%%%%%%
\begin{figure}[htb!]
\centering
\includegraphics[scale=0.7]{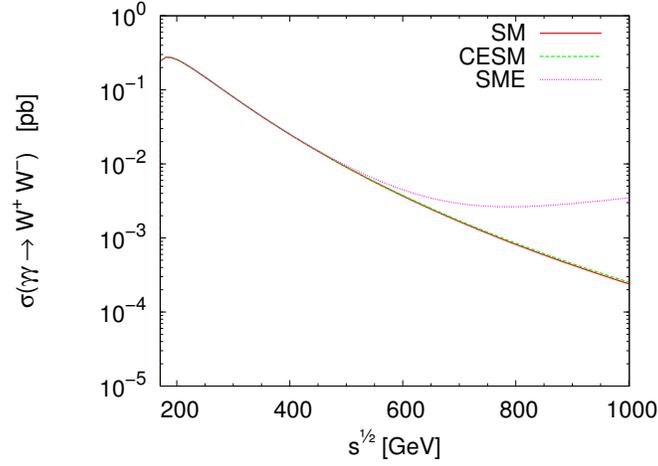}
\caption{Polarized cross section $(\pm,\pm, L, L)$ for the $\gamma\gamma\to WW$ process; integrated in the angular interval $20^\circ<\theta< 40^\circ$. Only the dominant scenario $\textbf{e}=0$, $\textbf{b}\neq 0$, with $\chi=176.96^\circ$, is considered.}
\label{SEppLL}
\end{figure}

In Fig.~\ref{SEppLL}, the polarized cross section $(\pm,\pm, L, L)$ is shown as a function of the center of mass energy. As it can be seen from this figure, when comparing the SM and CESM signals one can observe that they are practically the same. In addition, a clear Lorentz-violating effect arises from $\sqrt{s}\simeq600$ GeV, which is close to one order of magnitude larger than the SM contribution for $\sqrt{s}\simeq1000$ GeV.
%%%%%%%%%%%%%%%%%%%%%%%%%%%%%%%%%%%%%%%%%%%%%%%%%%%%%%%%%%%%%%%%%%%%%%%%%%%%%%%%%%%%%%%%%%%%%%
\begin{figure}[htb!]
\centering
\includegraphics[scale=0.7]{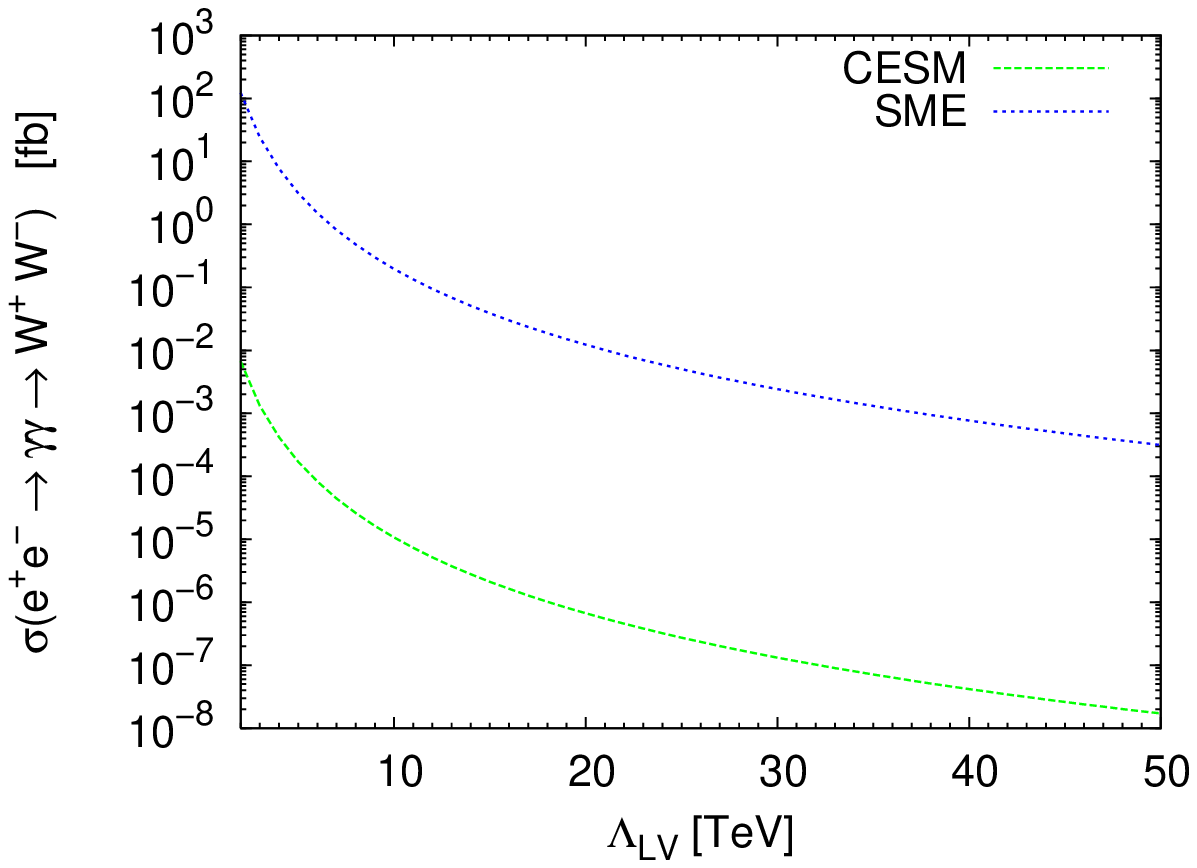}
\caption{Convoluted cross section $(\pm,\pm, (L,T+T,L))$ for the $\gamma\gamma\to WW$ process as a function of $\Lambda_{LV}$; integrated in the angular range $20^\circ<\theta< 40^\circ$. Only the dominant scenario $\textbf{e}=0$, $\textbf{b}\neq 0$, with $\chi=100^\circ$, is considered.}
\label{CCS}
\end{figure}
%%%%%%%%%%%%%%%%%%%%%%%%%%%%%%%%%%%%%%%%%%%%%%%%%%%%%%%%%%%%%%%%%%%%%%%%%%%%%%%%%%%%%%%%%%%%%%

Finally, we will carry out a study of the $W$ boson pair production in terms of the Lorentz-violating symmetry energy scale. The information collected will tell us whether it is feasible to study Lorentz-violating effects, consistent with Lorentz-violating energy scales obtained in recent studies~\cite{rlvbounds}. Surprisingly, the values founded disagree from each other, because the lower bounds ranges from $150$ GeV to $10^6$ TeV. Based on the above study, it is clear that the ideal scenario to study new physics effects, in particular, one where Lorentz symmetry is violated, corresponds to $\textbf{e}=0$, $\textbf{b}\neq 0$ with the $(\pm,\pm, (L,T+T,L))$ polarization, since this scenario offers pure contributions of new physics, i.e., there is no SM contribution. Thus, to analyze the possibility of detecting signals of Lorentz symmetry violation at the ILC through the $\gamma\gamma\to WW$ reaction it is essential to compute the convoluted cross section $e^+e^-\to\gamma\gamma\to WW$~\cite{Telnov}. Figure~\ref{CCS} shows the behavior of the convoluted cross section as a function of the LV energy scale which ranges from $2$ TeV to $50$ TeV. It is clearly seen that the CESM contribution is marginal when compared to the LV signal, since the latter one is 4 orders of magnitude larger throughout the energy scale interval considered. We now focus on discussing our final results within the experimental context of the ILC. At the last stage of operation, it is expected that this linear collider operates at the center of mass energy of $1$ TeV with an integrated luminosity of 1000 fb$^{-1}$~\cite{ILC-2013}. By considering this, we estimate that the number of events expected for the $(\pm,\pm,(L,T+T,L))$ polarization state are roughly 2 events for $\Lambda_{LV}=32$ TeV.

%%%%%%%%%%%%%%%%%%%%%%%%%%%%%%%%%%%%%%%%%%%%%%%%%%%%%%%%%%%%%%%%%%%%%%%%%%%%%%%%%%

\subsection{Asymmetries}
In Refs.~\cite{TSM1,RSM}, studies of various observables related with unpolarized and polarized total cross section were performed, due to their sensitivity to anomalous coupling effects, namely:
\begin{align}
R_{IO}&=\frac{\sigma(|\cos\theta|<0_{\cdot}4)}{\sigma(|\cos\theta|<0_{\cdot}8)},\\
R_{LT}&=\frac{\sigma_{LL}}{\sigma_{TT}},\\
R_{02}&=\frac{\sigma_{++}}{\sigma_{+-}}.
\end{align}
%%%%%%%%%%%%%%%%%%%%%%%%%%%%%%%%%%%%%%%%%%%%%%%%%%%%%%%%%%%%%%%%%%%%%%%%%%%%%%%%%%%%%%%%%%%%%%
\begin{figure}[htb!]
\centering
\includegraphics[scale=0.7]{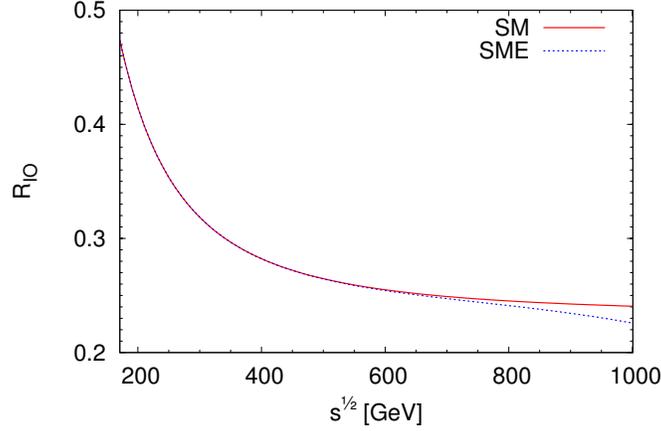}
\caption{$R_{IO}$ asymmetry for the $\gamma\gamma\rightarrow WW$ process. Only the dominant scenario $\textbf{e}=0$, $\textbf{b}\neq 0$ is considered.}
\label{RIO}
\end{figure}
%%%%%%%%%%%%%%%%%%%%%%%%%%%%%%%%%%%%%%%%%%%%%%%%%%%%%%%%%%%%%%%%%%%%%%%%%%%%%%%%%%%%%%%%%%%%%%
We are interested in studying the impact of Lorentz-violating symmetry effects on these asymmetries. Therefore, our analysis will focus on scenarios in which the SME contributions are enhanced ($\textbf{e}=0$, $\textbf{b}\neq 0$). In relation to the $R_{IO}$ asymmetry, the study of this observable will take place on the same scenario used to calculate the SM contribution~\cite{TSM1,RSM}. Regarding the remaining asymmetries, it is proposing a study scenario which comprises the angular region  $20^\circ<\theta< 40^\circ$ (LV cut), for $\chi=100^\circ$ ($\textbf{e}=0$, $\textbf{b}\neq 0$). For comparison reasons with previous works within the SM context~\cite{TSM1,RSM}, results with usual cut $|\cos\theta|=0.8$ will also be displayed.
%%%%%%%%%%%%%%%%%%%%%%%%%%%%%%%%%%%%%%%%%%%%%%%%%%%%%%%%%%%%%%%%%%%%%%%%%%%%%%%%%%%%%%%%%%%%%%
\begin{figure}[htb!]
\centering
\includegraphics[scale=0.7]{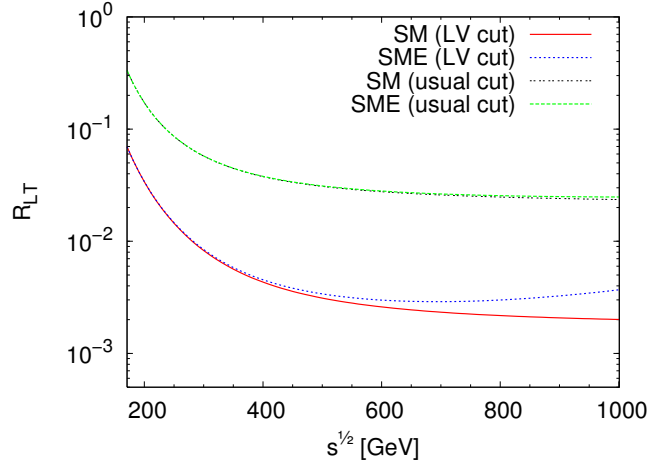}
\caption{$R_{LT}$ asymmetry for the $\gamma\gamma\rightarrow WW$ process. Only the dominant scenario $\textbf{e}=0$, $\textbf{b}\neq 0$ is considered.}
\label{RLT}
\end{figure}
%%%%%%%%%%%%%%%%%%%%%%%%%%%%%%%%%%%%%%%%%%%%%%%%%%%%%%%%%%%%%%%%%%%%%%%%%%%%%%%%%%%%%%%%%%%%%%

In Fig.~\ref{RIO}, the behavior of $R_{IO}$ as a function of the center of mass energy of the collision is shown. We have reproduced the SM lowest-order results given in Ref.~\cite{RSM}. It is important to add that this discussion does not include the CESM results, because its effects are suppressed and no relevant deviations from the SM contribution are observed. In this figure, from $\sqrt{s}\simeq800$ GeV to $\sqrt{s}=1$ TeV, it can be appreciated a clear deviation in the SME contribution from the SM signal, where the Lorentz-violating effect interferes negatively reducing the intensity of the SM asymmetry as the energy increases. From this figure, just at $\sqrt{s}=1$ TeV, we can observe a gap of $0.01$ in $R_{IO}$, which constitutes a signal of LV.

In Fig.~\ref{RLT}, the $R_{LT}$ asymmetry is displayed as a function of the center of mass energy for both usual cut and LV cut. In the former case, by using the angular region $|\cos\theta|=0.8$ we have reproduced the SM tree-level results~\cite{RSM}. Note that the CESM result is not presented since their new physics contributions are suppressed in the angular cut scenarios we are studying. Based on this figure, we see that when usual cut is imposed, no differences between the SME contribution and the SM signal are appreciated. However, when LV cut is imposed, the LV effect manifests itself in such a way that it increases significantly the SM contribution as the energy increases. Specifically, at $\sqrt{s}=$1 TeV there is a gap of approximately $2\times 10^{-3}$ unities between the two signals analyzed, which implies that a clear Lorentz-violating signal appears.

\begin{figure}[htb!]
\centering
\includegraphics[scale=0.7]{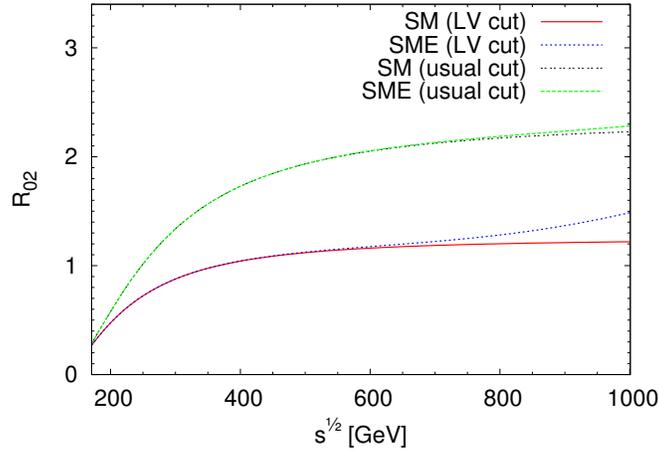}
\caption{Asimetría $R_{02}$ en la reacción $\gamma\gamma\rightarrow WW$ tomando un intervalo de $20^\circ<\theta< 40^\circ$, una cota de $|\cos\theta|<0_{\cdot}8$ y sin cota ($0^\circ<\theta< 180^\circ$), para SM y SME. Se considera sólo el escenario dominante ($\textbf{e}=0$, $\textbf{b}\neq 0$).}
\label{R02}
\end{figure}

In Fig.~\ref{R02}, we present the $R_{02}$ asymmetry as a function of the center of mass energy for both usual cut and LV cut. For the first case, the SM results were reproduced~\cite{RSM}. It should be noted that in the proposed study scenarios the CESM contribution is not shown since pure new physics contributions are suppressed. In the second case, from $\sqrt{s}\simeq800$ GeV to $\sqrt{s}=1$ TeV, it can be clearly observed a strong difference between both signals, where the LV effect splits up the SM prediction. At collision energies close to 1 TeV, it can be seen that pure new physics contributions leads to a strong deviation from the SM asymmetry, which is about of $0.3$ unities. Therefore, this effect represents a very clear signal of LV.
%%%%%%%%%%%%%%%%%%%%%%%%%%%%%%%%%%%%%%%%%%%%%%%%%%%%%%%%%%%%%%%%%%%%%%%%%%%%%%%%%%%%%%%%%%%%%%

\section{Final Remarks}
\label{C}
In this work, we have performed an exhaustive study of the helicity amplitudes for the $\gamma \gamma\to WW$ process, in which we have reproduced previous lowest-order results in the context of the SM, moreover, for the same reaction, we have computed tree-level order predictions in the context of both the SME and the CESM. The SME is a gauge-invariant extension of the SM, which incorporates Lorentz violation in a model-independent way, whereas the latter one is an effective theory which incorporates new physics effects, also in a model-independent fashion, but restricted to respect both the Lorentz and the SM gauge symmetries. An effective Yang-Mills Lagrangian for the electroweak sector of the SM was constructed through the introduction of a dimension-six Lorentz operator ${\cal O}_{\alpha \beta}$, which is invariant under the $SU_L(2)$ gauge group. New physics effects in the context of CESM were incorporated through the Lorentz invariant $g^{\alpha \beta}{\cal O}_{\alpha \beta}$. On the other hand, nonrenormalizable effects of Lorentz violation were considered in the context of the SME via the observer invariant $b^{\alpha \beta}{\cal O}_{\alpha \beta}$, with $b^{\alpha \beta}$ being an antisymmetric constant 2-tensor. The six components of this background constant tensor were parametrized in terms of the electric-like $\textbf{e}$ and magnetic-like $\textbf{b}$ spatial vectors. It was found that the best scenario in which the signal of Lorentz violation is free of the SM background and is high above the CESM contribution corresponds to $\textbf{e}=0$, $\textbf{b}\neq 0$ for the $(\pm,\pm,(L,T+T,L))$ state (there is no SM contribution), in the neighborhood of $\theta\approx20^\circ$ and  $\chi \approx 100^\circ$, where $\theta$ and $\chi$ are the scattering angle and the angle formed by the parallel component of $\textbf{b}$ with respect to collision line, respectively. The analysis of the differential cross section allowed us to conclude that the optimal angular region to study effects of Lorentz violation corresponds to $20^\circ<\theta<40^\circ$. For $|\textbf{b}|\sim 1$ and $\chi = 100^\circ$, the convoluted cross section for the state $(\pm,\pm,(L,T+T,L))$ on the interval $20^\circ<\theta<40^\circ$ leads to values which are between $10^2$ fb and $10^{-4}$ fb throughout the energy scale interval $2$ TeV$<\Lambda_{LV}<50$ TeV. We explored the possibility of detecting signals of Lorentz violation via the realistic process $e^+e^-\to\gamma\gamma\to WW$ at ILC. During the last stage of operation of this collider, it was estimated that with the projected integrated luminosity of $1000$ fb$^{-1}$ up to 2 events could be observed for a large energy scale of Lorentz-violating symmetry, $\Lambda_{LV}\approx32$ TeV.

\acknowledgments{We acknowledge financial support from CONACYT, CIC-UMSNH and
SNI (M\' exico).}

\appendix
\section{}\label{APA}
The explicit form of the $N^{b (i)}_{\alpha\beta\mu\nu\lambda\rho}$ structures is given below
\begin{align}
N^{b (1)}_{\alpha\beta\mu\nu\lambda\rho} =& (k^{\sigma}_{1}\delta^{\eta}_{\mu}-k^{\eta}_{1}\delta^{\sigma}_{\mu}) (k_{2\rho}\delta^{\chi}_{\nu}-k^{\chi}_{2}g_{\rho\nu}) \left(\frac{\Gamma^{1}_{\alpha\beta\sigma\eta\chi\lambda}}{2 k_{2} \cdot k_{4}}\right),\\
N^{b (3)}_{\alpha\beta\mu\nu\lambda\rho} =& (k^{\sigma}_{1}\delta^{\eta}_{\mu}-k^{\eta}_{1}\delta^{\sigma}_{\mu})  (k^{\chi}_{2}g_{\lambda\nu}-k_{2\lambda}\delta^{\chi}_{\nu}) \left(\frac{\Gamma^{3}_{\alpha\beta\sigma\eta\rho\chi}}{2 k_{2} \cdot k_{3}}\right),\\
N^{b (5)}_{\alpha\beta\mu\nu\lambda\rho} =& (k^{\sigma}_{1}\delta^{\eta}_{\mu}-k^{\eta}_{1}\delta^{\sigma}_{\mu}) \left(\frac{k_{4\nu}}{2 k_{2} \cdot k_{4}} - \frac{k_{3\nu}}{2 k_{2} \cdot k_{3}}\right) \Gamma^{5}_{\alpha\beta\sigma\eta\rho\lambda},\\
N^{b (7)}_{\alpha\beta\mu\nu\lambda\rho} =& (k^{\sigma}_{1}\delta^{\eta}_{\mu}-k^{\eta}_{1}\delta^{\sigma}_{\mu}) \left(\frac{k_{3\nu}}{2 k_{2} \cdot k_{3}}(k_{2\alpha}g_{\sigma\lambda}-k_{2\sigma}g_{\alpha\lambda}) - \frac{1}{2}(g_{\alpha\nu}g_{\sigma\lambda}-g_{\sigma\nu}g_{\alpha\lambda})\right) (k_{4\beta}g_{\eta\rho}-k_{4\eta}g_{\beta\rho}),\\
N^{b (11)}_{\alpha\beta\mu\nu\lambda\rho} =&(k^{\sigma}_{1}\delta^{\eta}_{\mu}-k^{\eta}_{1}\delta^{\sigma}_{\mu}) \left(\frac{k_{4\nu}}{2 k_{2} \cdot k_{4}} k_{3} \cdot k_{2} - \frac{1}{2} k_{3\nu}\right) (g_{\alpha\sigma}g_{\eta\rho}g_{\beta\lambda} - g_{\beta\sigma}g_{\alpha\rho}g_{\eta\lambda}), \\
N^{b (15)}_{\alpha\beta\mu\nu\lambda\rho} =& (k^{\sigma}_{1}\delta^{\eta}_{\mu}-k^{\eta}_{1}\delta^{\sigma}_{\mu}) \left(\frac{k_{4\nu}}{2 k_{2} \cdot k_{4}} k_{2\alpha} - \frac{1}{2} g_{\alpha\nu}\right) (k_{3\rho}g_{\eta\lambda} - k_{3\eta}g_{\lambda\rho})g_{\beta\sigma},\\
N^{b (19)}_{\alpha\beta\mu\nu\lambda\rho} =& (k^{\sigma}_{1}\delta^{\eta}_{\mu}-k^{\eta}_{1}\delta^{\sigma}_{\mu}) \left(\frac{k_{4\nu}}{2 k_{2} \cdot k_{4}} k_{2\eta} - \frac{1}{2} g_{\eta\nu}\right) (k_{3\beta}g_{\rho\lambda} - k_{3\rho}g_{\beta\lambda})g_{\alpha\sigma},\\
N^{b (23)}_{\alpha\beta\mu\nu\lambda\rho} =& (k^{\sigma}_{1}\delta^{\eta}_{\mu}-k^{\eta}_{1}\delta^{\sigma}_{\mu}) \left(\frac{k_{4\nu}}{2 k_{2} \cdot k_{4}} k_{2\lambda} - \frac{1}{2} g_{\lambda\nu}\right) (k_{3\eta}g_{\beta\sigma}g_{\alpha\rho} - k_{3\beta}g_{\alpha\sigma}g_{\eta\rho}),
\end{align}
with
\begin{align}
\Gamma^{1}_{\alpha\beta\sigma\eta\chi\lambda} =& (k_{3\sigma}g_{\alpha\lambda}-k_{3\alpha}g_{\sigma\lambda}) ((k_{3}-k_{1})_{\beta}g_{\eta\chi}-(k_{3}-k_{1})_{\eta}g_{\beta\chi}) \nonumber\\ & + g_{\beta\sigma}(k_{3\eta}\delta^{\omega}_{\lambda}-k^{\omega}_{3}g_{\eta\lambda}) ((k_{3}-k_{1})_{\alpha}g_{\omega\chi}-(k_{3}-k_{1})_{\omega}g_{\alpha\chi}) \nonumber\\ & + g_{\alpha\sigma}(k_{3\omega}g_{\beta\lambda}-k_{3\beta}g_{\omega\lambda}) ((k_{3}-k_{1})_{\eta}\delta^{\omega}_{\chi}-(k_{3}-k_{1})^{\omega}g_{\eta\chi}),\\
\Gamma^{3}_{\alpha\beta\sigma\eta\rho\chi} =& (k_{4\eta}g_{\beta\rho}-k_{4\beta}g_{\eta\rho}) ((k_{2}-k_{3})_{\alpha}g_{\sigma\chi}-(k_{2}-k_{3})_{\sigma}g_{\alpha\chi}) \nonumber\\ & + g_{\beta\sigma}(k_{4\omega}g{\alpha\rho}-k_{4\alpha}g_{\omega\rho}) ((k_{2}-k_{3})^{\omega}g_{\eta\chi}-(k_{2}-k_{3})_{\eta}\delta^{\omega}_{\chi}) \nonumber\\ & + g_{\alpha\sigma}(k^{\omega}_{4}g_{\eta\rho}-k_{4\eta}\delta^{\omega}_{\rho}) ((k_{2}-k_{3})_{\beta}g_{\omega\chi}-(k_{2}-k_{3})^{\omega}g_{\beta\chi}),\\
\Gamma^{5}_{\alpha\beta\sigma\eta\chi\lambda} =& (k_{3\sigma}g_{\alpha\lambda}-k_{3\alpha}g_{\sigma\lambda}) ((k_{4\eta}g_{\beta\rho}-k_{4\beta}g_{\eta\rho}) \nonumber\\ & + g_{\beta\sigma}(k_{3\eta}\delta^{\omega}_{\lambda}-k^{\omega}_{3}g_{\eta\lambda}) (k_{4\omega}g{\alpha\rho}-k_{4\alpha}g_{\omega\rho}) \nonumber\\ & + g_{\alpha\sigma}(k_{3\omega}g_{\beta\lambda}-k_{3\beta}g_{\omega\lambda}) (k^{\omega}_{4}g_{\eta\rho}-k_{4\eta}\delta^{\omega}_{\rho}.
\end{align}
The remaining gauge structures can be computed by the following exchanges:
\begin{align}
k_{1} \leftrightarrow k_{2}, \hspace{0.5cm} \mu\leftrightarrow\nu
 &\left\{\begin{array}{llllllll}
               &N^{b (1)}_{\alpha\beta\mu\nu\lambda\rho} \rightarrow N^{b (2)}_{\alpha\beta\mu\nu\lambda\rho} \\
               &N^{b (3)}_{\alpha\beta\mu\nu\lambda\rho} \rightarrow N^{b (4)}_{\alpha\beta\mu\nu\lambda\rho} \\
               &N^{b (5)}_{\alpha\beta\mu\nu\lambda\rho} \rightarrow N^{b (6)}_{\alpha\beta\mu\nu\lambda\rho} \\
               &N^{b (7)}_{\alpha\beta\mu\nu\lambda\rho} \rightarrow N^{b (8)}_{\alpha\beta\mu\nu\lambda\rho} \\
               &N^{b (11)}_{\alpha\beta\mu\nu\lambda\rho} \rightarrow N^{b (12)}_{\alpha\beta\mu\nu\lambda\rho} \\
               &N^{b (15)}_{\alpha\beta\mu\nu\lambda\rho} \rightarrow N^{b (16)}_{\alpha\beta\mu\nu\lambda\rho} \\
               &N^{b (19)}_{\alpha\beta\mu\nu\lambda\rho} \rightarrow N^{b (20)}_{\alpha\beta\mu\nu\lambda\rho} \\
               &N^{b (23)}_{\alpha\beta\mu\nu\lambda\rho} \rightarrow N^{b (23)}_{\alpha\beta\mu\nu\lambda\rho}
             \end{array},\right.\nonumber
\end{align}
\begin{align}
k_{3} \leftrightarrow k_{4}, \hspace{0.5cm} \lambda\leftrightarrow\rho & \left\{\begin{array}{lll}
               &N^{b (7)}_{\alpha\beta\mu\nu\lambda\rho} \rightarrow N^{b (9)}_{\alpha\beta\mu\nu\lambda\rho} \\
               &N^{b (11)}_{\alpha\beta\mu\nu\lambda\rho} \rightarrow N^{b (13)}_{\alpha\beta\mu\nu\lambda\rho}\\
               &N^{b (15)}_{\alpha\beta\mu\nu\lambda\rho} \rightarrow N^{b (17)}_{\alpha\beta\mu\nu\lambda\rho}\\
               &N^{b (19)}_{\alpha\beta\mu\nu\lambda\rho} \rightarrow N^{b (21)}_{\alpha\beta\mu\nu\lambda\rho}\\
               &N^{b (23)}_{\alpha\beta\mu\nu\lambda\rho} \rightarrow N^{b (25)}_{\alpha\beta\mu\nu\lambda\rho}
             \end{array},\right.\nonumber
\end{align}
\begin{align}
k_{1} \leftrightarrow k_{2}, \hspace{0.5cm} \mu\leftrightarrow\nu, \hspace{0.5cm} k_{3} \leftrightarrow k_{4}, \hspace{0.5cm} \lambda\leftrightarrow\rho
 &\left\{\begin{array}{lll}
               &N^{b (7)}_{\alpha\beta\mu\nu\lambda\rho} \rightarrow N^{b (10)}_{\alpha\beta\mu\nu\lambda\rho} \\
               &N^{b (11)}_{\alpha\beta\mu\nu\lambda\rho} \rightarrow N^{b (14)}_{\alpha\beta\mu\nu\lambda\rho}\\
               &N^{b (15)}_{\alpha\beta\mu\nu\lambda\rho} \rightarrow N^{b (18)}_{\alpha\beta\mu\nu\lambda\rho}\\
               &N^{b (19)}_{\alpha\beta\mu\nu\lambda\rho} \rightarrow N^{b (22)}_{\alpha\beta\mu\nu\lambda\rho}\\
               &N^{b (23)}_{\alpha\beta\mu\nu\lambda\rho} \rightarrow N^{b (26)}_{\alpha\beta\mu\nu\lambda\rho}
             \end{array}.\right.\nonumber
\end{align}

\section{}\label{APB}
The $E^{p,y}$ and $B^{p,y}$ expressions for the $W$ boson transversal components ($\lambda_{3,4}=\pm1$) are:
\begin{align}
E_{\lambda_1\lambda_2\lambda_3\lambda_4}^{p} =& \sin\theta\sin\phi\big\{\beta\big[16\big(\lambda_2(\lambda_4 - 2\lambda_3) +\lambda_1(\lambda_3 - 2\lambda_4)\big) - \big(\lambda_1(14\lambda_3 - \lambda_4) + \lambda_2(14\lambda_4 - \lambda_3)\big)\beta^2 \nonumber\\
& -\big(2\lambda_1 \lambda_2(1 - 16\lambda_3 \lambda_4) + 24\big)\beta\big] + 2\big[(4 - 6\lambda_1 \lambda_2 - 6\lambda_3 \lambda_4)\beta^3 + (2\lambda_1 \lambda_3 - 21(\lambda_2 \lambda_3 +\lambda_1 \lambda_4) \nonumber\\
& + 2\lambda_2 \lambda_4)\beta^2 - 2\big(14 - 4\lambda_3 \lambda_4 - \lambda_1 \lambda_2(3 - 16\lambda_3 \lambda_4)\big)\beta - 4(\lambda_1 + \lambda_2)(\lambda_3 + \lambda_4)\big]\cos\theta \nonumber\\
& + \beta\big[8\big((\lambda_1 + \lambda_2)(\lambda_4 + \lambda_3)\beta^2 - 4(\lambda_1 \lambda_2 \lambda_3 \lambda_4 +1)\beta + \lambda_1(\lambda_3 - 3\lambda_4) + \lambda_2(\lambda_4 - 3\lambda_3)\big)\cos(2\theta) \nonumber\\
& + 2\big(\beta(2\lambda_1 \lambda_3 - 7\lambda_2 \lambda_3 - 7\lambda_1 \lambda_4 + 2\lambda_2 \lambda_4 - 2\beta(\lambda_1 \lambda_2 + \lambda_3 \lambda_4 + 2)) - 2 (2 - \lambda_1 \lambda_2)\big)\cos(3\theta) \nonumber\\
& + \beta\big(-\beta\lambda_2(2\lambda_4 + \lambda_3) - \beta\lambda_1(2\lambda_3 + \lambda_4) + 2\lambda_1 \lambda_2 - 8\big)\cos(4\theta)\big]\big\}\cos\phi + 2\big\{2\big[-2(\lambda_1 \lambda_2 \nonumber\\
& + 5\lambda_3 \lambda_4 + 6)\beta^3 + (5\lambda_1 \lambda_3 - 4\lambda_2 \lambda_3 -4\lambda_1 \lambda_4 + 5\lambda_2 \lambda_4)\beta^2 + 2\big(6\lambda_3 \lambda_4 - \lambda_1 \lambda_2(8\lambda_3 \lambda_4 + 1) \nonumber\\
& - 2\big)\beta + 5(\lambda_1 + \lambda_2)(\lambda_3 + \lambda_4)\big] + \beta\big[\big(-\lambda_2 \lambda_3 \beta^2 - (10\lambda_1 \lambda_2 + 40)\beta + 2\lambda_3(11\lambda_1 - 7\lambda_2) \nonumber\\
& + (22\lambda_2 - \lambda_1(\beta^2 + 32\lambda_2 \lambda_3 \beta + 14))\lambda_4\big)\cos\theta + 2\big(-2(\lambda_1 \lambda_2 + \lambda_3 \lambda_4 +2)\beta^2 - 9(\lambda_2 \lambda_3 + \lambda_1 \lambda_4)\beta \nonumber\\
& + 2(\lambda_1 \lambda_2 - 2)\big)\cos(2\theta) + \beta\big(2\lambda_1(\lambda_2 - \beta\lambda_3) - \beta\lambda_2(\lambda_3 + 2\lambda_4) - \beta\lambda_1 \lambda_4 -8\big)\cos(3\theta)\big]\big\},
\end{align}
\begin{align}
E_{\lambda_1\lambda_2\lambda_3\lambda_4}^{y} =& -2\beta\big\{\big[(\lambda_2 - \lambda_1)(2\beta^2 -5) - (\lambda_1 \lambda_2 \lambda_3 + 2\lambda_3) \beta - \big(-\beta(\lambda_1 \lambda_2 + 2) \nonumber\\
& + 2(5\beta^2 -6)(\lambda_1 - \lambda_2)\lambda_3\big)\lambda_4\big]\sin\theta + 2\big[2(\lambda_4 - \lambda_3)\beta^2 - (\lambda_1 - \lambda_2)\lambda_3 \lambda_4 \beta \nonumber\\
& +\lambda_1 \lambda_2 (\lambda_4 - \lambda_3)\big]\sin(2\theta) + \big[2\lambda_2 (\lambda_3 \lambda_4 + 1)\beta^2 - 2\lambda_1 (\lambda_3 \lambda_4 + 1)\beta^2 \nonumber\\
& - \lambda_1 \lambda_2 (\lambda_3 - \lambda_4)\beta - 2(\lambda_3 - \lambda_4)\beta + \lambda_1 - \lambda_2\big]\sin(3\theta),
\end{align}
\begin{align}
B_{\lambda_1\lambda_2\lambda_3\lambda_4}^{p} =& -2\big\{\cos\theta\big[2\beta\big((\lambda_1 + \lambda_2)\beta^2 - (3 - 2\lambda_1 \lambda_2)(\lambda_3 + \lambda_4)\beta - 2(\lambda_1 + \lambda_2)(1 + 2\lambda_3 \lambda_4)\big) \nonumber\\
& - 4(\lambda_1 \lambda_2 + 1)(\lambda_3 + \lambda_4) + \beta\big((-5(\lambda_3 + \lambda_4)\beta^2 - (\lambda_1 + \lambda_2)(8\lambda_3 \lambda_4 + 5)\beta \nonumber\\
& + 4(\lambda_1 \lambda_2 + 1)(\lambda_3 + \lambda_4))\cos\theta + 2(-2\lambda_2 + \beta(\lambda_3 + \lambda_4 - \beta\lambda_2) - \lambda_1(\beta^2 + 4\lambda_2(\lambda_3 + \lambda_4)\beta \nonumber\\
& + 2))\cos(2\theta) + \beta(\beta\lambda_3 - 3\lambda_2 - \lambda_1(4\beta\lambda_2(\lambda_3 + \lambda_4) + 3) + \beta\lambda_4)\cos(3\theta)\big)\big]\cos\chi \nonumber\\
& + \big[2\big((\lambda_1 + \lambda_2)(3 + 2\lambda_3 \lambda_4)\beta^3 + ( 3 - 14\lambda_1 \lambda_2)(\lambda_3 + \lambda_4)\beta^2 - 2(\lambda_1 + \lambda_2)(2 - \lambda_3 \lambda_4)\beta \nonumber\\
& - 2(4\lambda_1 \lambda_2 + 3)(\lambda_3 + \lambda_4)\big) + \beta\big((3 - 32\lambda_1 \lambda_2)(\lambda_3 + \lambda_4)\beta^2 - (\lambda_1 + \lambda_2)(5 - 4\lambda_3 \lambda_4)\beta \nonumber\\
& - 4(4\lambda_1 \lambda_2 + 1)(\lambda_3 + \lambda_4)\big)\cos\theta +2\beta\big(-2\lambda_2 - \beta(\beta\lambda_2(1 + 2\lambda_3 \lambda_4) - 3(\lambda_3 + \lambda_4)) \nonumber\\
& - \lambda_1((2\lambda_3 \lambda_4 + 1)\beta^2 + 4\lambda_2(\lambda_3 + \lambda_4)\beta + 2)\big)\cos(2\theta) + \beta^2\big(\beta\lambda_3 - 3\lambda_2 \nonumber\\
& - \lambda_1(4\beta\lambda_2(\lambda_3 + \lambda_4) + 3) + \beta\lambda_4\big)\cos(3\theta)\big]\sin\theta\sin\chi\big\} ,
\end{align}
\begin{align}
B_{\lambda_1\lambda_2\lambda_3\lambda_4}^{y} =& \big[4(9\beta^2 + 1)\lambda_2 \lambda_3 + 2(15\beta^2 + 2)\lambda_2 \lambda_4 + \lambda_1\big(-(36\lambda_4 + 30\lambda_3)\beta^2 - 4(\lambda_3 + \lambda_4)\big)\big]\sin\theta \nonumber\\
& + \beta\big[(\lambda_2 + \lambda_1)(\lambda_4 - \lambda_3)\sin(4\theta)\beta^2 + 2\big((\lambda_2 - 4\lambda_1)\lambda_4 - (\lambda_1 - 4\lambda_2)\lambda_3\big)\sin(3\theta)\beta \nonumber\\
& - 2\big((\lambda_1 - \lambda_2)(10\lambda_3 \beta^2 + 10\lambda_4 \beta^2) + \lambda_1(7\lambda_4 - \lambda_3) - \lambda_2(7\lambda_3 - \lambda_4))\big)\sin(2\theta)\big].
\end{align}
The $E^{p,y}$ and $B^{p,y}$ functions for the $W$ boson longitudinal components ($\lambda^0_{3,4}=0$) are:
\begin{align}
E_{\lambda_1\lambda_2\lambda_3^0\lambda_4^0}^{p} =& 2\beta(\beta^2 - 1)\big[-\cos\theta\big(4\beta^2 + 5(4 - 3\lambda_1 \lambda_2)\beta\cos\theta + (4 - \lambda_1 \lambda_2)\beta\cos(3\theta) - 14\lambda_1 \lambda_2 \nonumber\\
& + (4 -4\beta^2 - 2\lambda_1 \lambda_2)\cos(2\theta) - 20\big)\cos\phi - \big(2(8\beta^2 - \lambda_1 \lambda_2 - 18) - \beta(3\lambda_1 \lambda_2 \nonumber\\
& + 28)\cos\theta + (4 - 8\beta^2 - 2\lambda_1 \lambda_2)\cos(2\theta) + \beta(4 - \lambda_1 \lambda_2)\cos(3\theta)\big)\sin\theta\sin\phi\big],\\
E_{\lambda_1\lambda_2\lambda_3^0\lambda_4^0}^{y} =& -2\beta(\beta^2 - 1)(\lambda_1 - \lambda_2)\big(-4\beta^2 + 10\beta\cos\theta + \cos(2\theta) + 13\big)\sin\theta,\\
B_{\lambda_1\lambda_2\lambda_3^0\lambda_4^0}^{p} =& \beta(\beta^2 - 1)(\lambda_1 + \lambda_2)\big\{\cos\theta\big[\beta\big(6\beta- 19\cos\theta + 3\cos(3\theta)\big) + 2(\beta^2 + 2)\cos(2\theta) \nonumber\\
& - 28\big]\cos\chi + \big[\beta\big(2\beta + \cos\theta + 3\cos(3\theta)\big) + 2(\beta^2 + 2)\cos(2\theta) - 4\big]\sin\theta\sin\chi\big\},\\
B_{\lambda_1\lambda_2\lambda_3^0\lambda_4^0}^{y} =& 0.
\end{align}
Finally, the $E^{p,y}$ and $B^{p,y}$ functions when are considered a transversal and a longitudinal components of the $W$ boson ($\lambda_{3}=\pm1$, $\lambda_{4}^0=0$) are:
\begin{align}
E_{\lambda_1\lambda_2\lambda_3\lambda_4^0}^{p} =& 2\big\{4\beta\big[\beta^2 - (3\beta^2 - 5)\lambda_1 \lambda_2\big] - 2\big[2(\beta^4 + \beta^2 + 1)\lambda_1 - (5\beta^4 - 11\beta^2 -2)\lambda_2\big]\lambda_3 \nonumber\\
& + \beta\big[\big((12 - 11\lambda_1 \lambda_2)\beta^3 - (22\lambda_1 - \lambda_2)\lambda_3 \beta^2 - 2(4 - 9\lambda_1 \lambda_2)\beta + 8(\lambda_1 - 3\lambda_2)\lambda_3\big)\cos\theta \nonumber\\
& + \big(4(\beta^2 + (1 - \beta^2)\lambda_1 \lambda_2 - 2) - 2\beta(4\lambda_1 \beta^2 - 3\lambda_2 \beta^2 - 2\lambda_1 - 7\lambda_2)\lambda_3\big)\cos(2\theta) \nonumber\\
& + \beta\big((\beta^2 - 2)(4 - \lambda_1 \lambda_2) - \beta(2\lambda_1 + \lambda_2)\lambda_3\big)\cos(3\theta)\big]\big\}\cos\phi\sin\theta + \big\{\beta\big((3\lambda_1 \lambda_2 + 20)\beta^3 \nonumber\\
& - (32\lambda_1 - 45\lambda_2)\lambda_3 \beta^2 - 6(4 - \lambda_1 \lambda_2)\beta + 2(23\lambda_1 - 27\lambda_2)\lambda_3\big) - 2\big((13\lambda_1 - 6\lambda_2)\lambda_3 \beta^4 \nonumber\\
& - 6(\lambda_1 \lambda_2 + 6)\beta^3 - (28\lambda_1 - \lambda_2)\lambda_3 \beta^2 + 2(18 - \lambda_1 \lambda_2)\beta + 10(\lambda_1 + \lambda_2)\big)\cos\theta \nonumber\\
& - \beta\big[2\big(-2(\lambda_1 \lambda_2 + 4)\beta^3 - (11\lambda_1 - 3\lambda_2)\lambda_3 \beta^2 + 2(4 - \lambda_1 \lambda_2)\beta + (11\lambda_1 - 7\lambda_2)\lambda_3\big)\cos(2\theta) \nonumber\\
& - 2\big(\beta(\beta^2 (3\lambda_1 - 4\lambda_2) + 9\lambda_2)\lambda_3 - 2(\beta^2 - 1)(2 - \lambda_1 \lambda_2)\big)\cos(3\theta) + \beta\big((\beta^2 - 2)(4 - \lambda_1 \lambda_2) \nonumber\\
& - \beta(\lambda_2 + 2\lambda_1)\lambda_3\big)\cos(4\theta)\big]\big\}\sin\phi,
\end{align}
\begin{align}
E_{\lambda_1\lambda_2\lambda_3\lambda_4^0}^{y} =& 2\big\{\big(7(\beta^2 - 3)\lambda_2 - 4\beta\lambda_3\big)\beta^2 + \big[-11\lambda_2 - \beta(3\lambda_3 \beta^2 - 4\lambda_2 \beta + 2\lambda_3) \nonumber\\
& + \lambda_1\big(17\beta((1 - \beta^2)\lambda_2 \lambda_3 - \beta) + 19\big)\big]\beta\cos\theta + \big(3(\lambda_2 - \lambda_1)\beta^3 - 2(3\lambda_1 \lambda_2 + 2)\lambda_3 \beta^2 \nonumber\\
& + (3\lambda_1 - 7\lambda_2)\beta - 2\lambda_1 \lambda_2 \lambda_3 \big)\beta\cos(2\theta) + \big[\lambda_1 - \lambda_2 - 3\lambda_1 \beta^2 - \big(\beta^2 + (1 - \beta^2)\lambda_1 \lambda_2 \nonumber\\
& + 2\big)\lambda_3 \beta\big]\beta\cos(3\theta) + 10\lambda_2 + \lambda_1\big[\beta\big(\beta(\beta^2 - 26\lambda_2 \lambda_3 \beta - 11) + 34\lambda_2 \lambda_3\big) + 10\big]\big\},
\end{align}
\begin{align}
B_{\lambda_1\lambda_2\lambda_3\lambda_4^0}^{p} =& -2\big\{-(6\lambda_2 +10\lambda_1)\beta^3 + 8(\lambda_1 + \lambda_2)\beta + \big[-\big((3\lambda_1 + 4\lambda_2)\beta^3 + 3(8\lambda_1 \lambda_2 + 7)\lambda_3 \beta^2 \nonumber\\
& + (5\lambda_1 - 7\lambda_2)\beta + 4\lambda_1 \lambda_2 \lambda_3 + 4\lambda_3\big)\cos\theta - 2\big((2 - \beta^2)\lambda_2 + \beta(5\beta^2 - 1)\lambda_3 \nonumber\\
& + \lambda_1(\beta^2 + 4\lambda_2 \lambda_3 \beta + 2)\big)\cos(2\theta) + \beta\big(\beta\lambda_3 - 3\lambda_2 + \lambda_1(3\beta^2 - 4\lambda_2 \lambda_3 \beta - 3)\big)\cos(3\theta)\big]\beta \nonumber\\
& + 2\big(3\beta^4 - (10\lambda_1 \lambda_2 + 9)\beta^2 - 2\lambda_1 \lambda_2 - 2\big)\lambda_3\big\}\cos\chi\sin\theta - \big\{(2\beta^4 - 29\beta^2 - 4)\lambda_2 \nonumber\\
& + \beta (21\beta^2 + 4)\lambda_3 + \lambda_1\big(-7\beta^4 + 23\beta^2 + 4(11\beta^2 + 4)\lambda_2 \lambda_3 \beta + 4\big) + \beta\big[2\big(-(3\lambda_2 + \lambda_1)\beta^3 \nonumber\\
& + 2(12\lambda_1 \lambda_2 + 5)\lambda_3 \beta^2 + 7(\lambda_1 - \lambda_2)\beta + 8\lambda_1 \lambda_2 \lambda_3 + 2\lambda_3\big)\cos(2\theta) + 2\big(-(\beta^2 + 2)\lambda_2 \nonumber\\
& + 3\beta(\beta^2 - 1)\lambda_3 + \lambda_1(\beta^2 + 4\lambda_2 \lambda_3 \beta + 2)\big)\cos(3\theta) - \beta\big(\beta\lambda_3 - 3\lambda_2 + \lambda_1(3\beta^2 - 4\lambda_2 \lambda_3 \beta \nonumber\\
& - 3)\big)\cos(4\theta)\big] + 2\big(\lambda_3 \beta^4 + (7\lambda_1 - 19\lambda_2)\beta^3 + 11(4\lambda_1 \lambda_2 + 1)\lambda_3 \beta^2 + 6(\lambda_1 - \lambda_2)\beta \nonumber\\
& + 4(3 + 4\lambda_1 \lambda_2)\lambda_3\big)\cos\theta\big\}\sin\chi,
\end{align}
\begin{align}
B_{\lambda_1\lambda_2\lambda_3\lambda_4^0}^{y} =& 2(-2\lambda_1 \lambda_2 + 7)\beta^4 + (17\lambda_1 - 11\lambda_2)\lambda_3 \beta^3 - 12(\lambda_1 \lambda_2 + 6)\beta^2 + 2(5\lambda_1 - \lambda_2)\lambda_3 \beta \nonumber\\
& + \big\{2\big[2\beta\big(-(\lambda_1 \lambda_2 + 4)\beta^2 - \lambda_1 \lambda_2 - 6\big) + \big((11\lambda_1 - 3\lambda_2)\beta^2 - \lambda_1 - 7\lambda_2\big)\lambda_3\big]\cos(2\theta) \nonumber\\
& + \beta\big[-2(3\beta\lambda_1 \lambda_2 + 4\lambda_3 \lambda_2 + \lambda_1 \lambda_2)\cos(3\theta) + \beta\big(2\beta + (\lambda_1 + \lambda_2)\lambda_3\big)\cos(4\theta)\big]\big\}\beta \nonumber\\
& - 8(3\lambda_1 \lambda_2 + 4) + 2\big(2(\lambda_1 + 3\lambda_2)\lambda_3 \beta^4 - (17\lambda_1 \lambda_2 + 48)\beta^3 + (19\lambda_1 - 16\lambda_2)\lambda_3 \beta^2 \nonumber\\
& - 4(\lambda_1 \lambda_2 + 4)\beta + 2(\lambda_1 - \lambda_2)\lambda_3\big)\cos\theta.
\end{align}
The remaining $E^{p,y}$ and $B^{p,y}$ functions ($\lambda_{3}^{0}=0$, $\lambda_{4}=\pm1$) are obtained from the above ones by replacing $\lambda_1 \leftrightarrow \lambda_2$ and $\lambda_3 \rightarrow \lambda_4$, which implies that
\begin{eqnarray}
E_{\lambda_1\lambda_2\lambda_3\lambda_4^0}^{p,y} &\rightarrow& E_{\lambda_1\lambda_2\lambda_3^0\lambda_4}^{p,y}, \nonumber\\
B_{\lambda_1\lambda_2\lambda_3\lambda_4^0}^{p,y} &\rightarrow& - B_{\lambda_1\lambda_2\lambda_3^0\lambda_4}^{p,y}. \nonumber
\end{eqnarray}

\end{document}